%% file: main.tex
\documentclass[sn-nature]{sn-jnl}% Style for submissions to Nature Portfolio journals

\input{_commands}

\input{_journal_abbrevs}
\input{_packages}

\begin{document}

\title{
    \textbf{
        Heavy element nucleosynthesis associated with a gamma-ray burst
    }
}

\author[1]{James~H.~Gillanders}
\email{jgillanders@roma2.infn.it;}
\affil[1]{\small{Department of Physics, University of Rome ``Tor Vergata'', via della Ricerca Scientifica 1, I-00133 Rome, Italy}}
\author[1]{Eleonora~Troja}
\email{eleonora.troja@uniroma2.it.}
\author[2,3,4,5,6]{Chris~L.~Fryer}
\affil[2]{\small{Computer, Computational, and Statistical Sciences Division, Los Alamos National Laboratory, Los Alamos, NM 87545, USA}}
\affil[3]{\small{Center for Theoretical Astrophysics, Los Alamos National Laboratory, Los Alamos, NM 87545, USA}}
\affil[4]{\small{The University of Arizona, Tucson, AZ 85721, USA}}
\affil[5]{\small{Department of Physics and Astronomy, The University of New Mexico, Albuquerque, NM 87131, USA}}
\affil[6]{\small{The George Washington University, Washington, DC 20052, USA}}
\author[7]{Marko~Risti\'c}
\affil[7]{\small{Center for Computational Relativity and Gravitation, Rochester Institute of Technology, Rochester, New York 14623, USA}}
\author[8,9,10]{Brendan~O'Connor}
\affil[8]{\small{Department of Physics, The George Washington University, 725 21st Street NW, Washington, DC 20052, USA}}
\affil[9]{\small{Department of Astronomy, University of Maryland, College Park, MD 20742-4111, USA}}
\affil[10]{\small{Astrophysics Science Division, NASA Goddard Space Flight Center, 8800 Greenbelt Rd, Greenbelt, MD 20771, USA}}
\author[3,11]{Christopher~J.~Fontes}
\affil[11]{\small{Computational Physics Division, Los Alamos National Laboratory, Los Alamos, NM, 87545, USA}}
\author[1]{Yu-Han~Yang}
\author[12]{Nanae~Domoto}
\author[12]{Salma~Rahmouni}
\author[12,13]{Masaomi~Tanaka}
\affil[12]{\small{Astronomical Institute, Tohoku University, Aoba, Sendai 980-8578, Japan}}
\affil[13]{\small{Division for the Establishment of Frontier Sciences, Organization for Advanced Studies, Tohoku University, Sendai 980-8577, Japan}}
\author[14]{Ori~D.~Fox}
\affil[14]{\small{Space Telescope Science Institute, 3700 San Martin Drive, Baltimore, MD 21218, USA}}
\author[15]{Simone~Dichiara}
\affil[15]{\small{Department of Astronomy and Astrophysics, The Pennsylvania State University, 525 Davey Lab, University Park, PA 16802, USA}}

\clearpage
\newpage

\abstract{
    \textbf{
        \noindent
        Kilonovae are a novel class of astrophysical transients, and the only observationally-confirmed site of rapid neutron capture nucleosynthesis (the \rpro) in the Universe \cite{Metzger2019, Cowan2021}. To date, only a handful of kilonovae have been detected \cite{Troja2023}, with just a single spectroscopically-observed event (\gfo; \cite{Abbott2017}). Spectra of \gfo\ \cite{Pian2017, Smartt2017} provided evidence for the formation of elements heavier than iron \cite{Watson2019}; however, these spectra  were collected during the first $\sim 10$ days, when emission from light \rpro\ elements dominates the observations. Heavier elements, if synthesised, are expected to shape the late-time evolution of the kilonova \cite{Wu2019, Kasliwal2022}, beyond the phases for which we have spectral observations. Here we present spectroscopic observations of a rapidly-reddening thermal transient, following the gamma-ray burst, \thisGRB. Early (2.4~day) optical spectroscopy identifies the presence of a hot ($T \approx 6700$\,K) thermal continuum. By 29~days, this component has expanded and cooled significantly \mbox{($T \approx 640$\,K)}, yet it remains optically thick, indicating the presence of high-opacity ejecta. We show that these properties can only be explained by the merger of compact objects, and further, leads us to infer the production of the heavy lanthanide elements. We identify several spectral features (in both absorption and emission), whose cause can be explained by newly-synthesised heavy elements. This event marks only the second recorded spectroscopic evidence for the synthesis of \rpro\ elements, and the first to be observed at such late times.
    }
}

\maketitle

\section*{Main text}

The gamma-ray burst (GRB) 230307A was discovered on 7 March 2023 at 15:44:07~UTC (hereafter $T_0$) by NASA's \textit{Fermi} satellite \cite{Fermi_GRB_detection2023GCN.33405, Fermi_GRB_detection2023GCN.33407, Fermi_GRB_detection2023GCN.33411}, and later accurately localised by NASA's \textit{Neil Gehrels Swift Observatory} \cite{Swift_GRB_followup2023GCN.33419, Swift_GRB_followup2023GCN.33429}. Early observations across the entire electromagnetic (EM) spectrum detected a bright non-thermal transient that followed the GRB and rapidly faded. This emission, typically referred to as afterglow \cite{Costa1997}, arises from the interaction of the relativistic outflow with its surrounding medium \cite{Meszaros1997, Wijers1997}. Multi-colour imaging of the source led to the identification of an additional component, in excess of the standard afterglow, at phases $\gtrsim 1$ day \yuhancite. This component, modelled as a thermal transient, is seen to rapidly shift toward infrared (IR) wavelengths.

\begin{figure}
    \centering
    \includegraphics[width=0.5\linewidth]{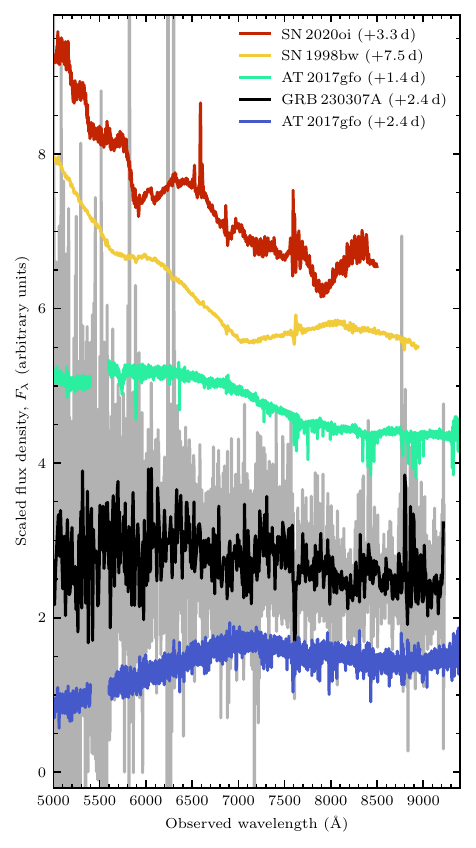}
    \caption{
        \textbf{Gemini spectrum of \thisGRB, compared to early-time spectra of SNe and the KN, \gfo.}
        After subtracting the contribution of the non-thermal afterglow from the optical spectrum, we identify a featureless thermal continuum at $T_0 + 2.4$~days, resembling the shape of a KN, rather than a SN. The original spectrum is plotted (grey), with the $3 \sigma$-clipped and $5 \times$ re-binned version over-plotted (black). The \gfo\ spectra have been $3 \sigma$-clipped and $5 \times$ re-binned for clarity, while the spectra of SNe 1998bw and 2020oi are unaltered. The spectra are vertically offset for clarity.
    }
    \label{fig: Gemini vs KNe and GRB-SNe}
\end{figure}

In addition to the comprehensive photometric dataset \yuhancite, several spectra of the GRB counterpart were acquired. We utilised the available spectroscopic dataset to constrain the GRB's distance scale, and identify the origin of its rapidly-evolving thermal emission. First, we obtained an optical spectrum of the GRB counterpart with the 8.1\,m Gemini South telescope at a phase of $T_0 + 2.4$ days (Figure~\ref{fig: Gemini vs KNe and GRB-SNe}). There is no clear evidence in this spectrum for any commonly-observed absorption or emission features. Specifically, the spectrum contains no evidence for the Lyman limit, responsible for suppressing emission blueward of $912 (1 + z)$\,\AA. The high-significance detection of continuum emission redward of $\sim 5000$\,\AA\ (observer frame) places an upper limit to the GRB redshift, $z \lesssim 4.5$. Additional constraints can be derived from the lack of any other prominent spectral features, such as \lyalpha\ absorption. This absorption feature is prominent in GRB afterglow spectra \cite{Selsing2019} (Extended Data Figure~\ref{fig: GRBs 130408A + 230307A spectra}). The lack of this feature in the optical spectrum of \thisGRB\ strongly indicates that \lyalpha\ lies blueward of the observed continuum emission ($\lesssim 5000$\,\AA), implying a more stringent constraint to the redshift, $z \lesssim 3.1$ (Methods).

Additional spectra were obtained by the \textit{James Webb Space Telescope} (\jwst), at epochs of $T_0 + 29$ and $T_0 + 61$ days (Extended Data Figure~\ref{fig: JWST spectra - Epochs 1+2}; \cite{Levan2023arXiv}). These spectra sample near- to mid-infrared (NIR--MIR) wavelengths, with coverage spanning $\approx 0.6 - 5.3$\,\micron. Both \jwst\ spectra display a prominent narrow emission line at $\approx 32060$\,\AA. This emission line arises from a faint galaxy, offset from \thisGRB\ by $\sim 0.24$ arcsec \cite{Levan2023arXiv, Yang_partner_paper}. If produced by H$\alpha$, this emission line corresponds to a redshift, $z \approx 3.89$. As H$\alpha$ is the reddest of the most commonly-observed emission lines, this value sets a robust lower limit to the galaxy's distance scale. Not only are there tensions between this high redshift and the observed high-energy properties of this burst \cite{Yang_partner_paper, Sun2023} but, as discussed above, our Gemini spectrum possesses no evidence in support of a high-redshift origin. We therefore conclude that the angular proximity of \thisGRB\ to this \mbox{high-$z$} galaxy is a chance alignment. Following the detailed analysis of \yuhancite, we place \thisGRB\ at redshift, \mbox{$z = 0.0647 \pm 0.0003$}, by assuming the explosion occurred at the same distance of its most probable host galaxy. This corresponds to a luminosity distance, $D_{\textsc{l}} \approx 291$\,Mpc (assuming $\Lambda$CDM cosmology with a Hubble constant, \mbox{$H_0 = 69.8$\,\kmsMpc}, \mbox{$\Omega_{\rm M} = 0.315$} and \mbox{$\Omega_\Lambda = 0.685$}; \cite{Freedman2019, PlanckCollaboration2020}), thus making \thisGRB\ one of the closest GRBs ever detected.

Having established the distance to \thisGRB, we now turn to a physical interpretation of its unusual thermal emission. We find evidence for emission in excess of the expected afterglow contribution at $T_0 + 2.4$~days (Extended Data Figure~\ref{fig: Spectra versus AG+BB fits}, Methods). Initially, we modelled the spectrum with a single power-law function to represent the expected spectral energy distribution (SED) of the non-thermal afterglow emission. While a single power-law with exponent $\beta_\lambda = - 2.25 \pm 0.07$ can reproduce the optical spectrum, this derived slope significantly overestimates the X-ray flux at this phase. We interpret this as evidence of a thermal component contributing to the optical band. This interpretation is further corroborated by the spectral shape at later times. Both \jwst\ spectra are inconsistent with the expected afterglow contribution (Extended Data Figure~\ref{fig: Spectra versus AG+BB fits}). The first \jwst\ spectrum ($T_0 + 29$ days) exhibits a very red colour with a steep continuum ($\beta_\lambda \approx + 1.02 \pm 0.02$) consistent with the blue end of a blackbody component. The second \jwst\ spectrum ($T_0 + 61$ days) exhibits marked evolution compared to the first, confirming the transient nature of the thermal component. 

To investigate the properties of this component, we modelled the spectra with the combination of a power-law continuum and a blackbody profile (Methods). This allows us to isolate the afterglow contribution and estimate the photospheric temperature ($T_{\rm ph}$) and radius ($R_{\rm ph}$) of the thermal component. Assuming homologous expansion then leads to an estimate for the photospheric velocity ($v_{\rm ph}$). The best-fitting models are presented in Extended Data Figure~\ref{fig: Spectra versus AG+BB fits}, with associated model parameters presented in Extended Data Table~\ref{tab: BB temperatures and velocities}. Based on this analysis, we identify several spectral features at $T_0 + 29$~days (Methods). These can be interpreted as emission-like features rising above the observed continuum, with peak wavelengths of $\sim 2.1$ and 4.4\,\micron, as well as an absorption-like feature suppressed below the continuum, centred at \mbox{$\sim 3.4$\,\micron} (Figure~\ref{fig: JWST +29d spec, BB fits, continuum subtractions + Gaussian features}). We find the $\sim 2.1$\,\micron\ emission feature can be readily explained by either a single component, or a blend of two components. The location of this feature is comparable to a prominent emission feature visible in the late-time spectra of \gfo\ \cite{Gillanders2023arXiv} (Extended Data Figure~\ref{fig: JWST spectra - Epochs 1+2}).

An alternative continuum fit recovers the same emission feature at $\sim 2.1$\,\micron\ and absorption feature at $\sim 3.4$\,\micron, but indicates the presence of another absorption-like feature, centred at $\sim 4.0$\,\micron\ (Figure~\ref{fig: JWST +29d spec, BB fits, continuum subtractions + Gaussian features}). We note that this prominent $\sim 2.1$\,\micron\ emission feature is also present in the later \jwst\ spectrum, although it has faded by a factor of $\sim 5$, and shifted redward by $\sim 800$\,\AA\ (Extended Data Table~\ref{tab: Gaussian velocities}). From this empirical analysis, we conclude the following: (i)~the ejecta exhibits a high expansion velocity, $v_{\rm ph} \sim 0.1 \, c$, and rapid cooling, from $T_{\rm ph} \approx 6700$\,K at $T_0 + 2.4$~days to \mbox{$T_{\rm ph} \approx 640$\,K} at \mbox{$T_0 + 29$~days}; (ii)~there is evidence for optically thick material at least up to \mbox{$T_0 + 29$~days}; (iii)~spectral features are present at $T_0 + 29$ and $T_0+ 61$~days (in emission and absorption). We use these observational constraints to probe the ejecta properties and identify the nature of the explosion.

The inferred high expansion velocities match previous observations of GRB supernovae (GRB-SNe; \eg, \cite{Taddia2019}) and kilonovae (KNe; \eg, \cite{Drout2017, Pian2017, Smartt2017}), both of which are transient phenomena previously associated with GRBs. In Figure~\ref{fig: Gemini vs KNe and GRB-SNe}, we compare the optical spectrum of \GRBxx{230307A} with early spectra obtained for the GRB-SN 1998bw \cite{Patat2001}, the type Ic SN 2020oi \cite{Gagliano2022}, and the KN, \gfo\ \cite{Pian2017, Smartt2017}.\footnote{Spectra for SNe 1998bw and 2020oi were obtained from the WISeREP repository (\url{https://wiserep.weizmann.ac.il}; \cite{Yaron2012_WISeREP}). We utilise the flux-calibrated spectra of \gfo, publicly available on the ENGRAVE webpages (\url{https://www.engrave-eso.org/AT2017gfo-Data-Release}).} Both SN spectra display a blue color and prominent absorption features, neither of which match the Gemini spectrum of \thisGRB. The overall spectral shape of \thisGRB\ best matches the \gfo\ spectrum obtained at +1.4~days. The evolution across all three epochs of spectral observations of \thisGRB\ is too rapid to match the evolutionary timescales of high-mass ejecta ($\sim$~a few solar masses) from typical GRB-SNe, but does match expectations for the low-mass ejecta ($\lesssim$~a tenth of a solar mass) from KNe and other compact object mergers. We therefore consider the possibility that the thermal emission associated with \thisGRB\ is produced by a compact object merger, involving a white dwarf (WD) or neutron star (NS), merging with a NS or stellar mass black hole (BH). These mergers \mbox{(WD--NS/BH, NS--NS/BH)} produce rapidly-evolving transients (\eg, \cite{Drout2017, Gillanders2020, Yang2022}), as observed in the case of \thisGRB.

We begin by examining the case of a KN from a NS--NS/BH merger. KNe are powered by the radioactive decay of many rapidly-synthesised heavy isotopes \cite{Li1998, Metzger2019, Korobkin2021} produced via the rapid neutron capture process (\rpro; \cite{Cowan2021}). A KN naturally accounts for the luminosity, red colour and rapid evolution of the transient; however, the late-time properties are unprecedented for a KN, and not readily explained by current quantitative models. There have been several previous identifications of KN events associated with GRBs \cite{Troja2023}, but only one of these has been observed \mbox{\textit{spectroscopically} -- \gfo} \cite{Pian2017, Smartt2017}. \gfo\ exhibited rapid evolution, transitioning from a blue to a red kilonova after just $\sim 2 - 3$ days \cite{Drout2017}, and exhibiting deviations from a photospheric spectrum after $\sim$~1~week \cite{Gillanders2021}. Given the low ejecta mass and rapid expansion velocity inferred for \gfo\ (\mbox{$M_{\rm ej} \approx 0.05$\,\msun} and $v_{\rm ej} \approx 0.1 \, c$; \cite{Waxman2018}), the ejecta transitions away from an optically thick regime (\ie, $\tau \gtrsim 1$) for opacities, $\kappa < 0.5 \left( \frac{t}{7 \, {\rm d}} \right)^2$\,\cmg.

In contrast to these expectations, the first \jwst\ spectrum of \thisGRB\ appears to be dominated by a red thermal continuum. Sophisticated two-component KN models, which include radiative transfer and detailed opacity information \cite{Wollaeger2018, Korobkin2021, Ristic2022}, predict a spectrum rich in emission features that does not match the observed spectral shape for any plausible range of ejecta mass and velocity (Methods). This mismatch occurs due to the model opacity $\gtrsim 2$\,\micron\ being too low to reproduce the red continuum.

Given this limitation, we adopt a simplified semi-analytic model, which allows us to empirically constrain the basic properties of the ejecta (\eg, mass, velocity, and opacity) without complete and detailed atomic line information \cite{Metzger2019}. This model contains no radiative transfer, and assumes a one-zone, single component ejecta, with total mass $m$ (split into shells with mass $m_v$, distributed as a function of velocity $v$), minimum velocity $v_0$, wavelength-independent (\ie, grey) opacity $\kappa$, and a power-law parameter $\beta$ which describes the mass-distribution of the ejecta (Methods). After subtracting the contribution of the non-thermal afterglow, our best-fitting model can reproduce the \jwst\ spectrum at $T_0 + 29$~days by invoking \mbox{$m = 0.059^{+0.010}_{-0.012}$}\,\msun, \mbox{$v = 0.088^{+0.008}_{-0.013} \, c$} and \mbox{$\kappa = 10^{+9}_{-5}$\,\cmg} (Extended Data Figures~\ref{fig: metzger_fits}~and~\ref{fig: metzger_corner}). This analysis confirms that an optically thick spectrum at such late times requires high opacities -- a statement that applies to both WD--NS/BH and NS--NS/BH mergers.

\begin{figure}
    \centering
    \begin{subfigure}[b]{0.495\linewidth}
        \begin{overpic}[width=\linewidth]{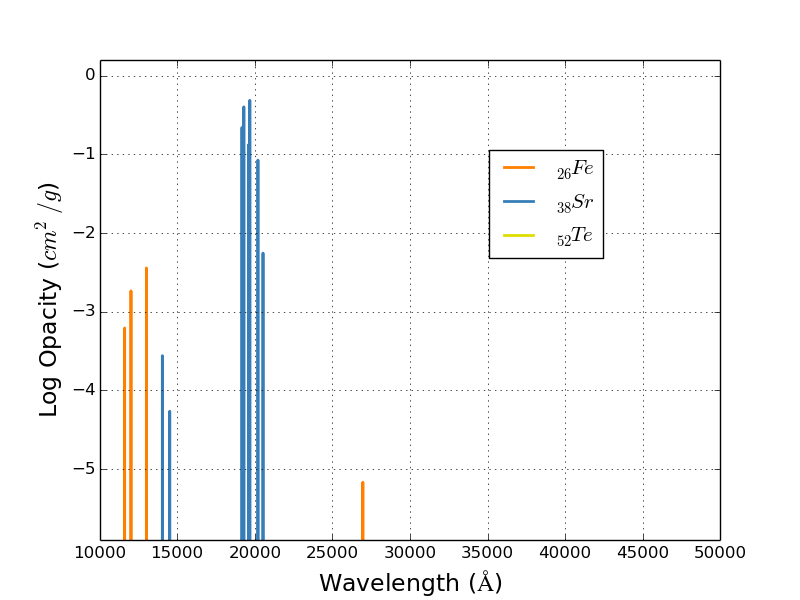}
            \put (14,63) {\footnotesize \textbf{A}}
        \end{overpic}
    \end{subfigure}
    \begin{subfigure}[b]{0.495\linewidth}
        \begin{overpic}[width=\linewidth]{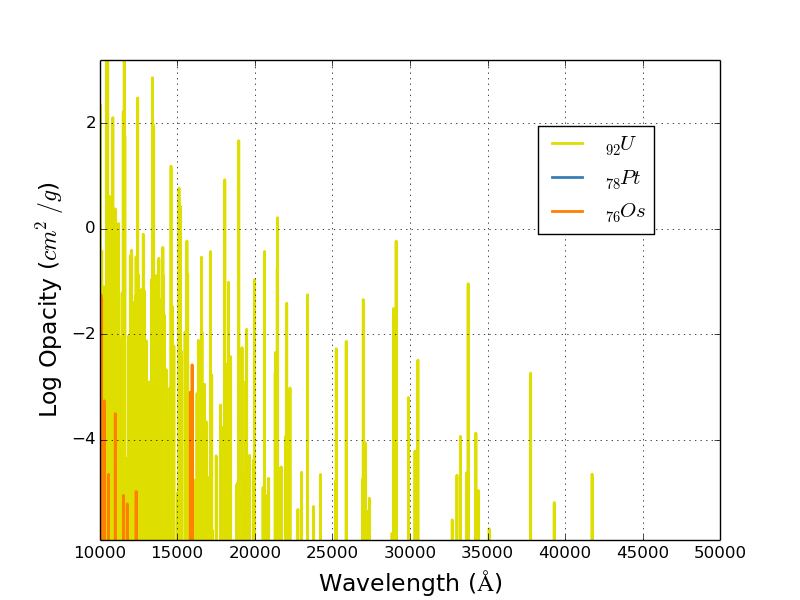}
            \put (14,63) {\footnotesize \textbf{B}}
        \end{overpic}
    \end{subfigure}
    \begin{subfigure}[b]{0.495\linewidth}
        \begin{overpic}[width=\linewidth]{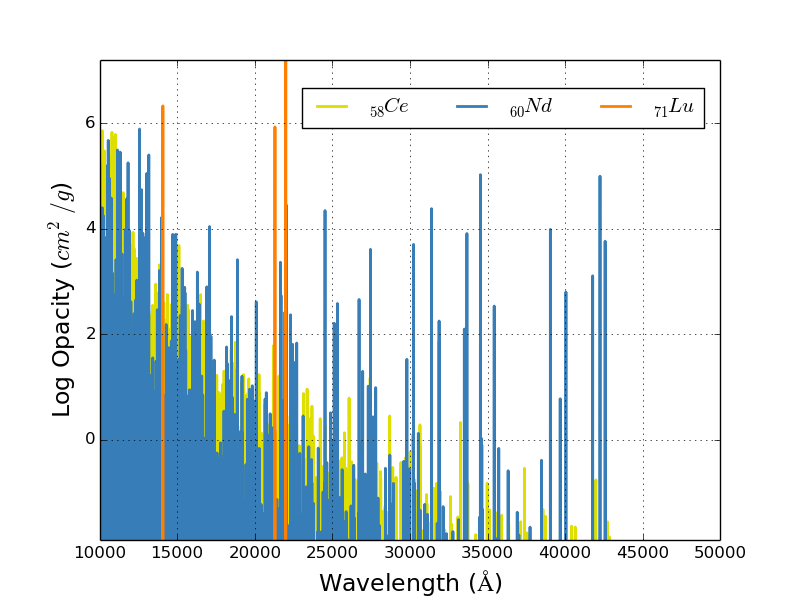}
            \put (14,63) {\footnotesize \textbf{C}}
        \end{overpic}
    \end{subfigure}
    \caption{
        \textbf{Opacity versus wavelength for a range of characteristic \rpro\ elements.}
        We assume LTE conditions, with a temperature, $T = 0.07$\,eV, and density, $\rho = 10^{-16}$\,\gcm. Note the opacity scale (y-axis) varies considerably for each panel.  These opacities visualise the numbers of lines for different elements in the wavelength range of the \jwst\ observations.  For iron and the first and second \rpro\ peaks, there are very few lines, and we expect the opacities from these elements to be \mbox{$\ll 0.01\, {\rm cm}^2 \, {\rm g}^{-1}$}. Some actinides, such as U, possess many more lines, but the opacity $\gtrsim 2.5$\,\micron\ is still $\lesssim 0.01\, {\rm cm}^2 \, {\rm g}^{-1}$.
        \textbf{Panel~A:} Opacities for iron (Fe), strontium (Sr; first \rpro\ peak) and tellurium (Te; second \rpro\ peak).
        \textbf{Panel~B:} Opacities for osmium and platinum (Os and Pt, respectively; third \rpro\ peak) and uranium (U; actinide).
        \textbf{Panel~C:} Opacities for cerium, neodymium and lutetium (Ce, Nd and Lu, respectively; lanthanides).
    }
    \label{fig: opacfin}
\end{figure}

We now explore two possible scenarios that may explain the inferred high opacities at these wavelengths and at this phase -- heavy elements in the case of a NS--NS/BH merger, and molecules in the case of a WD--NS/BH merger. In KN models, atomic processes (\ie, bound--bound transitions) dominate the opacity. Under the inferred conditions at $T_0 + 29$~days \mbox{($T_{\rm ph} \approx 600 - 650$\,K)}, it is already difficult to produce the required continuum opacities above $\sim 2 - 3$\,\micron. This issue is exacerbated further if one also attributes the MIR emission at $T_0 + 61$~days to a thermal continuum (Extended Data Figure~\ref{fig: Spectra versus AG+BB fits}). Assuming local thermodynamic equilibrium (LTE), iron-peak elements and first \rpro\ peak elements possess some line features $< 3$\,\micron, but nothing that will produce strong lines $> 3$\,\micron. Only heavy elements, such as the lanthanides (\eg, cerium, neodymium), have strong lines $> 3$\,\micron\ (Figure~\ref{fig: opacfin}). With sufficient Doppler broadening and the inclusion of more detailed opacity, these lines can potentially provide the `continuum-like' opacities required to produce a photosphere.

At these low temperatures, non-thermal effects are expected to be significant, and, as we shall discuss below, may be required to explain the observed emission features. However, we note that lines from non-thermal effects (\eg, forbidden transitions) are not able to produce the dense forest of lines needed to generate the high continuum opacities. We conclude that for KN models, lanthanides (or other heavy \rpro\ elements; \eg, actinides) must be present in the ejecta to explain the existence of a photosphere (Methods).

We further investigate which \rpro\ species are able to explain the spectral features present in emission. Following Ref.~\cite{Gillanders2023arXiv} (see also Supplementary Material), we shortlist a number of forbidden transitions that are coincident with the two emission features (Extended Data Table~\ref{tab: Forbidden candidate transitions}). Of particular interest is tellurium (Te). We recover the same [Te\,\III] transition proposed by \cite{Levan2023arXiv} for the feature at $\sim 2.1$\,\micron, in addition to finding agreement with a prominent [Te\,\II] transition and the $\sim 4.4$\,\micron\ feature. Tellurium lies at the top of the second \rpro\ abundance peak, and, therefore, it is expected to be abundant in KN ejecta across many different \rpro\ scenarios \cite{Gillanders2022, Gillanders2023arXiv}. We also shortlist multiple ions of the lanthanides neodymium and erbium (Nd and Er, respectively), which are capable of matching the observed emission. These are expected to be among the most abundant lanthanide elements synthesised \cite{Goriely2011, Goriely2013, Goriely2015, Bauswein2013, Gillanders2022}, and their presence is supported by the high opacity of the inner ejecta, and by the rapid evolution of the photosphere \yuhancite. Therefore, invoking them as the cause of these spectral features seems plausible. Singly-ionised platinum (Pt\,\II), a third \rpro\ peak element, may also contribute to the emission feature at $\sim 2.1$\,\micron\ \cite{Gillanders2022}.

The possible source of the two absorption features, centered at \mbox{$\sim 3.4$ and $4.0$\,\micron} (Figure~\ref{fig: JWST +29d spec, BB fits, continuum subtractions + Gaussian features}), is less certain. We identify a number of strong candidate transitions for doubly-ionised cerium and thorium (Ce\,\III\ and Th\,\III) which tend to produce strong absorption lines at NIR and MIR wavelengths \cite{Domoto2022}. However, they both present some inconsistencies with the observed data (Supplementary Material). Regardless of the specific identification, the spectral features of \thisGRB\ can only be explained by KN models if robust \rpro\ nucleosynthesis took place, in agreement with the conclusions derived from the opacity calculations.

If instead the transient associated with \thisGRB\ was produced by a carbon--oxygen WD--NS merger, the slowest-moving ejecta is expected to be predominantly carbon~(C) and oxygen~(O) \cite{Kaltenborn2022arXiv}. Although the atomic opacities of these elements would be low, the inferred photospheric temperature supports contribution from both molecular and dust opacities. Both CO molecular and dust features have been observed in a number of SNe $100 - 300$\,d after explosion \cite{Tinyanont2023}. CO molecular features are expected to arise at temperatures $\lesssim 3000$\,K. Because the WD--NS merger models tend to have ejecta velocities that are $\sim 3 - 10 \times$ higher than normal SNe, adiabatic expansion (where temperature is inversely proportional to both velocity and time) predicts that this ejecta will achieve similar conditions $\sim 3 - 10 \times$ earlier than normal SNe. As expected, our inferred ejecta temperatures at $\sim 30$ and 60~days are similar to those of SNe at 300\,d (\eg, \cite{Tinyanont2023}).  The combined molecular and dust opacities may explain the continuum opacity required to produce the photosphere (Methods) and, in addition, provide a match to the observed line features.

The line identifications for the features presented in Figure~\ref{fig: JWST +29d spec, BB fits, continuum subtractions + Gaussian features} are very different in the \mbox{WD--NS/BH} merger scenario. WD--NS/BH mergers eject large amounts of C, O and calcium (Ca). Assuming LTE opacities, Ref.~\cite{Kaltenborn2022arXiv} predict strong Ca\,\I\ lines \mbox{$\sim 2$\,\micron}, which could contribute to the $\sim 2.1$\,\micron\ emission feature of \thisGRB. Additionally, non-thermal effects could produce prominent Ca\,\II\ features at 2.1395 and 2.1435\,\micron. At these temperatures, the CO molecular line features should also be present, with first overtone features spanning $1.95 - 2.35$\,\micron, and second overtone features spanning $4.1 - 5.0$\,\micron\ (assuming a bulk blueshift of $0.1 \, c$), potentially explaining both the emission features (see Extended Data Figure~\ref{fig: JWST +29d continuum subtracted spec with WD--NS/BH line IDs}). We note the relative strengths of these features matches expectations for the strengths of the first and second overtones, with the $\sim 2.1$\,\micron\ feature being more luminous than the $\sim 4.4$\,\micron\ feature.

Based on the analysis of the late-time \jwst\ spectra, we cannot break the degeneracy between the two progenitor models, as they both provide a plausible interpretation of the observational features. However, whereas a KN from a NS--NS/BH merger matches the featureless Gemini spectrum at $T_0 + 2.4$~days, the transient from a WD--NS/BH merger is expected to be rich in absorption features \cite{Gillanders2020, Kaltenborn2022arXiv}, seemingly at odds with the early spectral observations. The identification of \thisGRB\ as a NS--NS/BH merger is also supported by its rapid reddening, with its spectrum peaking in the $K$-band at $\sim T_0+7$~days \yuhancite. Based on our current understanding of WD--NS/BH mergers and their electromagnetic counterparts, we consider these systems a less likely progenitor for \thisGRB. NS--NS/BH mergers provide a more consistent explanation of the full dataset, and are backed by observational evidence from past KNe. The rate of events is also consistent with the range inferred from \textit{Swift} observations ($0.04 - 0.8$\,Gpc$^{-3}$\,yr$^{-1}$\,sky$^{-1}$; \cite{Troja2022}).

The proposed classification as a NS--NS/BH merger means that \thisGRB\ doubles the sample of spectroscopically-observed KNe, while also providing the first spectral observations of a KN at late times ($\sim 61$~days post-explosion) and at MIR wavelengths ($\lambda \leq 5$\,\micron). Whereas the identification of individual \rpro\ species remains uncertain, the presence of a thermal continuum is a strong indicator of high-opacity lanthanide element nucleosynthesis.

%End of main text -- now for main text Figures and Tables.
\clearpage
\newpage

\begin{figure}
    \centering
    \includegraphics[width=\linewidth]{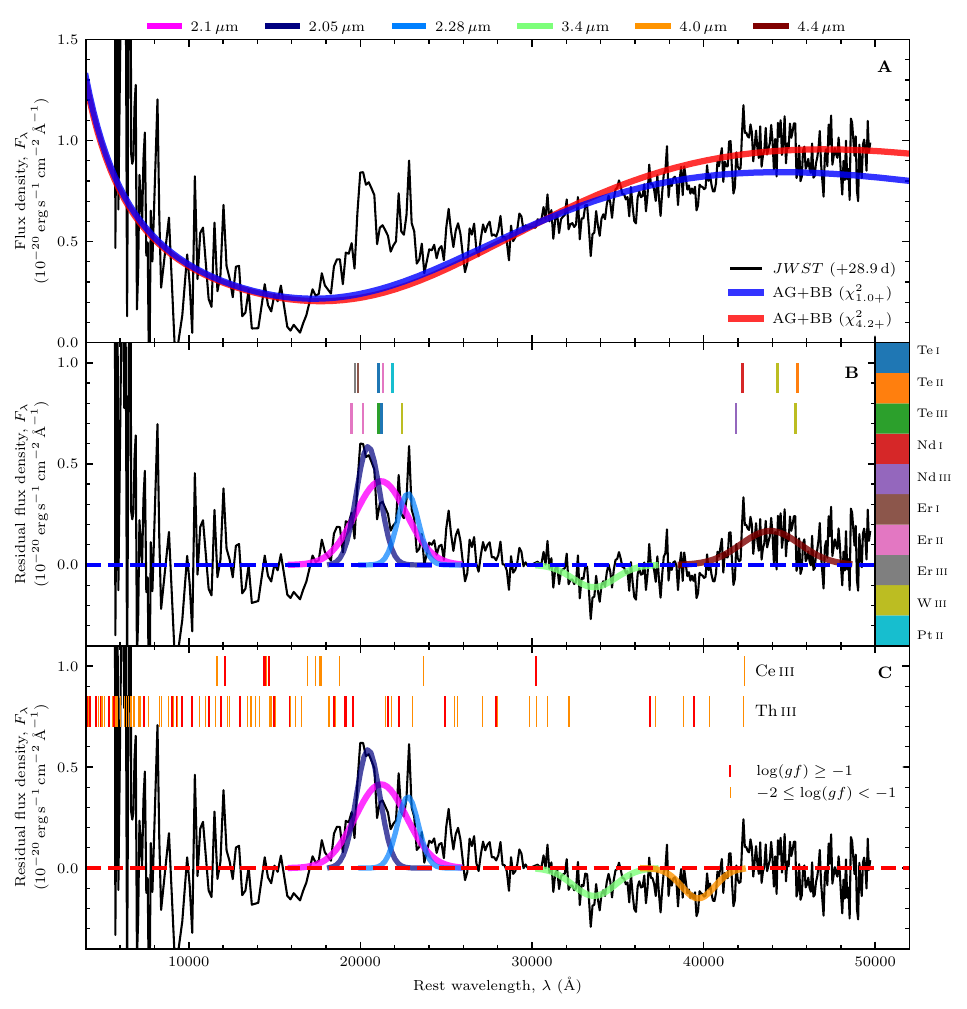}
    \caption{
        \textbf{Identification of spectral features in the $T_0 + 29$~day \jwst\ spectrum of \thisGRB.}
        \textbf{Panel~A:} Comparison between the \jwst\ spectrum and our best-fit combined afterglow and blackbody (AG+BB) profiles -- both the \chisqall\ and \chisqred\ models (Methods). The observed spectrum has been corrected for extinction along the line of sight \cite{Schlafly2011}, and shifted to the rest frame. The emission line from a background galaxy has been masked.
        \textbf{Panel~B:} Identification of emission and absorption features in the observed spectrum, obtained after subtracting the \chisqall\ model continuum. Some of the proposed line identifications for emission are displayed (see Extended Data Table~\ref{tab: Forbidden candidate transitions} for the full list).
        \textbf{Panel~C:} Identification of emission and absorption features in the \jwst\ spectrum, obtained after subtracting the \chisqred\ model continuum. The strongest Ce\,\III\ and Th\,\III\ transitions with low-lying levels that are expected to possess a large Sobolev optical depth are plotted, with a blueshift, $v = 0.1 \, c$ (see Supplementary Material).
    }
    \label{fig: JWST +29d spec, BB fits, continuum subtractions + Gaussian features}
\end{figure}

% Start of the Methods section
\clearpage
\newpage

\section*{Methods}

\section{Distance constraints}

Our Gemini/GMOS spectrum offers the most direct constraint to the distance scale of \thisGRB, which is a key parameter for its physical interpretation. The lack of absorption and emission features prevents an accurate redshift measurement, and also inform us that the GRB is not located in a highly star-forming region. The most stringent limit that we can place is derived from the absence of \lyalpha\ absorption, which is a prominent feature in GRB afterglow spectra. The legacy sample of \xsh\ GRB afterglows \cite{Selsing2019} reports the identification of \lyalpha\ in all 42 GRBs located at $z> 2$, where the feature is visible by the spectrograph. 

Gemini/GMOS has a narrower wavelength coverage, but is still sensitive to the \lyalpha\ feature for $z \gtrsim 3$. As an example, we show the Gemini South spectroscopy of \GRBxx{130408A} \cite{GRB130408A_Gemini_redshift_2013GCN.14366} at $z \approx 3.757$ (Extended Data Figure~\ref{fig: GRBs 130408A + 230307A spectra}). The spectrum was reduced using standard recipes in \dragons\ \cite{DRAGONS_OG_paper, DRAGONS_v3.0.4_zenodo}, and calibrated using observations of a standard star. Around the time of the observations, \GRBxx{130408A} was \mbox{$r' \approx 19.7 \pm 0.1$ AB mag} \cite{Sudilovsky2013GCN.14364}, 8 times brighter than \thisGRB\ (\mbox{$r = 22.0 \pm 0.3$ AB mag}; \yuhancite). However, the reduced spectrum shown in Extended Data Figure~\ref{fig: GRBs 130408A + 230307A spectra} consists of a 600\,s exposure, $\sim 7$ times shorter than the total integration time for \thisGRB. Therefore, the sensitivity of the reduced spectra should be somewhat comparable.

As shown in Extended Data Figure~\ref{fig: GRBs 130408A + 230307A spectra}, there are prominent absorption features in the spectrum of \GRBxx{130408A}, arising from \lyalpha, \lybeta\ and the Lyman limit, in addition to various lines belonging to C\,\II, O\,\I\ and S\,\II. Clearly, none of these features are present in the spectrum of \thisGRB. An equivalently-prominent \lyalpha\ absorption feature should be observable at $\sim 5950$\,\AA, if \thisGRB\ is located at $z = 3.89$, ruling out a physical association with the high-$z$ background galaxy visible in the \jwst\ observations. Moreover, since we do not see this feature at any position $\gtrsim 5000$\,\AA\ in our observed spectrum, we can estimate a more constraining upper limit for \thisGRB\ to be $z \lesssim 3.1$. 

\section{Identification of an additional emission component}

GRBs are followed by a long-lived non-thermal afterglow that, over a narrow energy band, is typically well described by a simple power-law spectrum: \mbox{$F_{\lambda} \propto \lambda^{\beta_\lambda}$}, where $F_\lambda$ is the flux density (in $\lambda$-space) and $\beta_\lambda$ is the spectral index. Note that since we present all spectra in this manuscript in terms of $F_\lambda$, we opt to present the afterglow modelling in $F_\lambda$ space too. Our afterglow model parameters are easily converted into $F_\nu$ space, with the exponents for both related by: \mbox{$\beta_\lambda = - ( 2 + \beta_\nu )$}. 

The Gemini spectrum is reasonably reproduced by a single power-law, where \mbox{$F_{\lambda} = (1.4 \pm 0.8) \times 10^{-9} \, \lambda^{-2.25 \pm 0.07}$\,\ergscmA}. However, this fit significantly over-predicts the flux at high energies \yuhancite, and is not consistent with the spectral index derived from X-ray spectroscopy and broadband SED modelling ($\beta_\lambda = - 1.20 \pm 0.02$; \yuhancite). With this additional constraint, we quantitatively estimate the afterglow's flux contribution to the optical spectrum. Comparing our estimated value to the observed data shows evidence for an emission component in addition to the standard afterglow (Extended Data Figure~\ref{fig: Spectra versus AG+BB fits}). 

The initial \jwst\ ($T_0 + 29$~day) spectrum has a steep, red continuum, and clearly demonstrates an excess compared to the expected afterglow contribution. We again use the X-ray data \yuhancite\ as a tracer of the non-thermal emission, and estimate its contribution to longer wavelengths, adopting $\beta_\lambda = - 1.37_{-0.05}^{+0.07}$ \yuhancite. We find that the afterglow dominates the observed spectrum $\lesssim 2$\,\micron, but cannot produce the flux redward of this wavelength (see Extended Data Figure~\ref{fig: Spectra versus AG+BB fits}). The later \jwst\ spectrum ($T_0 + 61$~days) is significantly fainter (see Extended Data Figure~\ref{fig: JWST spectra - Epochs 1+2}). However, there is still a detectable signal, allowing us to further explore any excess to the afterglow at this phase. Using the same methodology and spectral index, we again find evidence for a flux excess (see Extended Data Figure~\ref{fig: Spectra versus AG+BB fits}).

\section{Empirical modelling} \label{sec:Methods - Thermal component modelling}

We model the emission in excess of the standard afterglow with a Planck function, characterised by two free parameters -- the effective temperature and radius of the emitting surface ($T_{\rm ph}$ and $R_{\rm ph}$, respectively). This choice is motivated by the steep rising continuum observed by \jwst, consistent with a blackbody profile, and by the analysis of the broadband SED \yuhancite. The best-fit model is obtained by minimising the residual $\chi^2$ values between the model and the observed data. From the best-fit parameters, we derive the expansion velocity of the ejecta (assuming homologous expansion, \ie, \mbox{$R_{\rm ph} = v_{\rm ph} \times t$}, where $t$ is the time since explosion) and the luminosity of the transient, $L = 4 \pi R_{\rm ph}^2 \sigma_{\textsc{sb}} T_{\rm ph}^4$, where $\sigma_{\textsc{sb}}$ is the Stefan-Boltzmann constant.

The properties of the best-fit thermal component for each spectral epoch are listed in Extended Data Table~\ref{tab: BB temperatures and velocities}. For the \jwst\ spectrum at $T_0 + 29$~days, we report two possible models. The former, dubbed `\chisqall', is obtained by fitting the observed spectrum for wavelengths, $\lambda \geq 1.0$\,\micron\ (with the prominent emission feature at $\sim 2.1$\,\micron\ and the background galaxy line masked). Above 4.2\,\micron, we identify emission that lies systematically above the model continuum (Figure \ref{fig: JWST +29d spec, BB fits, continuum subtractions + Gaussian features}). We therefore present an alternative description of the thermal component, by interpreting the flux redward of $4.2$\,\micron\ as the true continuum level, as opposed to an excess from an additional source. This model, dubbed `\chisqred', is obtained by fitting the observed spectrum for $\lambda \geq 4.2$\,\micron. This fit leads to minor variations in the inferred properties of the thermal continuum but, as described below, it affects the identification of the spectral features.

\section{Spectral features}

From our empirical fit (\chisqall) of the first \jwst\ spectrum, we identify three broad deviations from our model continuum. These appear to be two emission-like features rising above the observed continuum, with peak wavelengths of $\sim 2.1$ and 4.4\,\micron, and an absorption-like feature at $\sim 3.4$\,\micron\ (see Figure~\ref{fig: JWST +29d spec, BB fits, continuum subtractions + Gaussian features}). Alternatively, based on the \chisqred\ model, we identify two absorption-like features, centred at $\sim 3.4$ and 4.0\,\micron, in addition to the prominent $\sim 2.1$\,\micron\ emission-like feature. The $\sim 2.1$\,\micron\ feature is also visible in the late-time \jwst\ spectrum. 

In Figure~\ref{fig: JWST +29d spec, BB fits, continuum subtractions + Gaussian features} we compare both model continua (\chisqall and \chisqred) to the observed \jwst\ spectrum. We also present the continuum-subtracted spectra obtained from both interpretations, and highlight the locations of the spectral features in emission and in absorption. We fit these features with Gaussian profiles, and extract estimates for their peak wavelength, full-width, half-maximum (FWHM) velocity ($v_{\textsc{fwhm}}$) and total flux -- summarised in Extended Data Table~\ref{tab: Gaussian velocities}. From this fitting procedure, we derive $v_{\textsc{fwhm}} \sim 0.06 - 0.18 \, c$, comparable with the inferred photospheric velocities. 

The presence of these broad spectral absorption and emission features is indicative of the existence of some ion(s) in the ejecta of \thisGRB\ influencing the observed continuum shape. As such, we performed a line identification study to constrain the presence of emitting species in the ejecta (as detailed in Ref.~\cite{Gillanders2023arXiv} and in the Supplementary Material). Our search returned matches with light (\eg, silver, tellurium), intermediate (\eg, neodymium, erbium) and heavy (\eg, tungsten, platinum) \rpro\ elements that possess transitions in coincidence with the observed emission features (see Extended Data Table~\ref{tab: Forbidden candidate transitions} for the full list). We searched for matches to the absorption features with Ce\,\III\ and Th\,\III\ (as in Ref.~\cite{Domoto2022}), but found no convincing explanation for these ions dominating these spectral features (see Supplementary Material).

\section{Kilonova modelling with \supernu}

We explore whether the observed thermal emission could be interpreted as a kilonova. To this aim, we apply the Los Alamos National Laboratory (LANL) radiative transfer code, \supernu\ \cite{supernu}, to model the spectrum at $T_0 + 29$~days. \supernu\ lightcurves and spectra have proven successful in recovering estimates for the ejecta properties of \gfo\ \cite{Evans2017, Troja2017, Ristic2022, Ristic2023arXiv, Kedia2023}, motivating its application here.

We use simulated two-component (lanthanide-rich dynamical ejecta and lanthanide-poor wind ejecta) spectra derived from the LANL simulation library \cite{Wollaeger2018, Wollaeger2021, Korobkin2021, Ristic2022}. Due to the prevalence of the signal in the infrared, we expect the dynamical ejecta component to dominate. The higher opacities in this ejecta, largely due to the atomic line contributions of heavy \rpro\ elements, lead to the re-processing of photons to longer wavelengths, not observed in the wind component. Given the detection of a GRB, we only consider spectra for nearly on-axis viewing angles ($\theta \lesssim 16$\degree).

Using the same approach detailed in Ref.~\cite{Ristic2023arXiv}, we infer a dynamical ejecta mass \mbox{$m_d = 0.0625^{+0.0002}_{-0.0044}$\,\msun}, and velocity \mbox{$v_d = 0.1228^{+0.0002}_{-0.0010} \, c$}. As we discuss below, the small errors in these inferred models are, in part, an artifact of the model assumptions. Extended Data Figure~\ref{fig: fit_lanl_m19_sigmaobs} compares the $T_0 + 29$~day \jwst\ spectrum with the simulated spectrum, corresponding to the best-fitting \supernu\ model parameters. The main discrepancy between the model and observation is in the continuum trend -- the observed spectrum trends upward for longer wavelengths, whereas the simulated spectrum demonstrates the opposite behavior, trending downward, with a sharp fall-off $\lesssim 3.8$\,\micron. There is some minor agreement at $\sim 4.3$\,\micron\ where the model can match the observed spectrum, but beyond $\sim 3.8$\,\micron, the flux is typically too under-luminous to fit the observed red spectrum.

In addition, we also explored a range of low-mass ($< 0.001 - 0.005$\,\msun) ejecta models with different compositions. In these \supernu\ models, the photosphere has receded significantly, and most of the emission arises from optically-thin regions, where electrons are still sufficiently trapped to thermalise and produce emission. If the opacity was dominated by iron-peak or first \rpro\ peak elements, the emission would be much bluer than what is observed. Models with heavy \rpro\ distributions match the observed level of IR flux; however, their spectra display strong spectral features rather than a smooth continuum. Either our models are missing sufficient atomic data information to fill in these spectral features (we do not use a full complement of the lanthanide and actinide elements in these models), or a true photosphere is necessary to interpret the red continuum observed by \jwst. 

\section{Kilonova semi-analytic modelling}

Motivated by our inability to reproduce the observed spectra with \supernu\ models, we shift our modelling of the thermal continuum to an empirically-motivated analysis using the semi-analytic model presented by \cite{Metzger2019} (hereafter referred to as \metzger).

The \metzger\ model assumes a one-zone, single-component ejecta with total mass $m$ (split into mass shells $m_v$ which are distributed as a function of velocity $v$), minimum velocity $v_0$, wavelength-independent (grey) opacity $\kappa$, and a power-law parameter $\beta$ which describes the mass-distribution of the ejecta in velocity space, as $m_v = m(\frac{v}{v_0})^{-\beta}$. The model calculates the energy evolution of the system which is assumed to be expanding homologously (\ie, non-accelerating) in the radial direction. In our case, the energy evolution focusses only on heating by \rpro\ radioactive ($\beta^-$) decays and cooling via radiative losses and ejecta expansion. The \rpro\ heating prescription uses the analytic heating rate presented in \cite{Korobkin2012}, with thermalisation efficiencies from \cite{Barnes2016}. Radiative emission is measured from the photosphere, identified as the radius at which the optical depth, $\tau \approx 1$. The final model output produces the spectral energy density $F_\lambda$ emitted by a blackbody with photospheric temperature and radius, $T_{\rm ph}$ and $R_{\rm ph}$. Throughout this calculation, the model does not perform any radiative transfer.

Spectra are generated using iterative grids of input model parameters drawn from our prior-constrained parameter space (\mbox{$-3 \leq \log [m_{\rm{ej}} \, (\msun)] \leq -1$}, \mbox{$0.01 \leq v \, (c) \leq 0.30$}, \mbox{$-1 \leq \log [\kappa \, ({\rm cm}^2 \, {\rm g}^{-1})] \leq 1.477$}, \mbox{$1 \leq \beta \leq 8$}). The final posteriors presented in Extended Data Figure~\ref{fig: metzger_corner} are generated using $3 \times 10^6$ total samples. Agreement to the observed data is calculated using reduced $\chi^2$ statistics. The best-fit spectrum compared to the $T_0 + 29$~day \jwst\ spectrum is shown in Extended Data Figure~\ref{fig: metzger_fits}. The best-fit model parameters and their 1$\sigma$ confidence intervals are \mbox{$m = 0.059^{+0.010}_{-0.012}$}\,\msun, \mbox{$v = 0.088^{+0.008}_{-0.013} \, c$}, and \mbox{$\kappa = 10^{+9}_{-5}$\,\cmg}. The velocity distribution parameter $\beta$ is largely unconstrained; however, its effects on the fit are minor compared to those of the opacity and mass. The main result derived from this analysis is that high opacities are required to reproduce the observed thermal continuum at infrared wavelengths. 

As cautioned above, the \metzger\ model makes several simplifying assumptions; however, these do not affect our inference about high opacities. In this model, the ejecta mass is primarily set by the energy deposition needed to produce the observed luminosities. The one-zone model is unable to capture many effects of radiation transport that can alter our final solutions. For example, in full transport models, radiation (both thermal and non-thermal photons) from the inner ejecta can heat the material at the photosphere. This additional energy source would decrease the required mass in our fits. The trend from this effect would be for our current set of models to over-estimate the ejecta mass, and under-estimate the opacity. 

\section{White dwarf -- neutron star mergers}

The long-duration of the gamma-ray emission could suggest a less compact progenitor system, such as a WD--NS/BH merger \cite{Fryer1999, Kaltenborn2022arXiv, Zhong2023}. Current WD--NS merger models produce a wide range of bolometric light curves \cite{Bobrick2022, Kaltenborn2022arXiv}, some of which are consistent with the observed luminosity of \thisGRB. Due to the fast $^{56}$Ni ejecta, their spectra are initially blue, but become redder over time \cite{Kaltenborn2022arXiv}. Other models are instead always dominated by their reddest band (\eg, $y$-band; \cite{Bobrick2022}), although none of these models or simulations achieve the rapid reddening observed for \thisGRB.

Moreover, given that the heaviest elements in these mergers are typically limited to the iron-peak, achieving the high atomic opacities required to produce a photosphere at late-times is difficult (Figure~\ref{fig: opacfin}). CO molecules dominate the slow-moving ejecta in these WD--NS merger models. At temperatures $\lesssim 1000$\,K, CO molecular lines have been observed in SNe \cite{Tinyanont2019, Rho2021}, and these present a good match to the observed emission features in \thisGRB, assuming a modest $\sim 0.1 \, c$ bulk blueshift (see Extended Data Figure~\ref{fig: JWST +29d continuum subtracted spec with WD--NS/BH line IDs}). Molecular features can also contribute to the inferred high infrared opacities. 

Our models of WD--NS mergers match the current state-of-the-art simulations using \supernu\ (or similar high-order transport schemes), assuming LTE opacities. Extended Data Figure~\ref{fig: WDNSmer} shows the simulated spectra at 30\,d for a range of WD--NS merger models, whose ejecta are sensitive both to the masses of the WD and NS, and the evolution of the disk and its wind \cite{Kaltenborn2022arXiv}. Because the opacities are derived assuming LTE, Ca\,\I\ lines produce a strong feature $\approx 2$\,\micron\ \cite{Kaltenborn2022arXiv}, that could plausibly explain the observed feature in \thisGRB.

\section{Ejecta opacity}

The high opacities required to produce a photosphere in the infrared are evidence that lanthanides (or other heavy \rpro\ elements, \eg, actinides) exist in the ejecta. Figure~\ref{fig: opacfin} shows the binned LTE opacities from the LANL database \cite{NIST-LANL}, assuming a density, $\rho = 10^{-16}$\,\gcm. At the low radiation temperatures expected in our kilonova models ($T \lesssim 1000$\,K), the atoms are expected to be in their neutral states (according to LTE), and only lanthanides and actinides (\eg, Nd, U) have strong features. It has been argued that high-energy electrons produced via $\beta$-decay can drive the atoms to higher ionisation states than expected by collisions from thermal electrons \cite{Hotokezaka2021}. To try to understand the spectra from atoms at higher ionisation states, we plot the Fe and Te spectra at 3\,eV, when these elements are doubly-ionised. Even with these conditions, we cannot produce the high opacities needed to explain the photosphere and match the observed red spectra. Therefore, to obtain high atomic opacities beyond $\sim 1 - 2$\,\micron, heavier \rpro\ species (\ie, lanthanides or actinides) are required. We have varied both the temperature and density to determine the sensitivity of these results on the exact conditions and, although specific line features vary, these basic conclusions do not change.

% Start of Extended Data Figures and Tables
\clearpage
\newpage

\renewcommand{\figurename}{Extended Data Figure}
\renewcommand{\tablename}{Extended Data Table}
\setcounter{figure}{0}

\begin{figure}
    \centering
    \includegraphics[width=\linewidth]{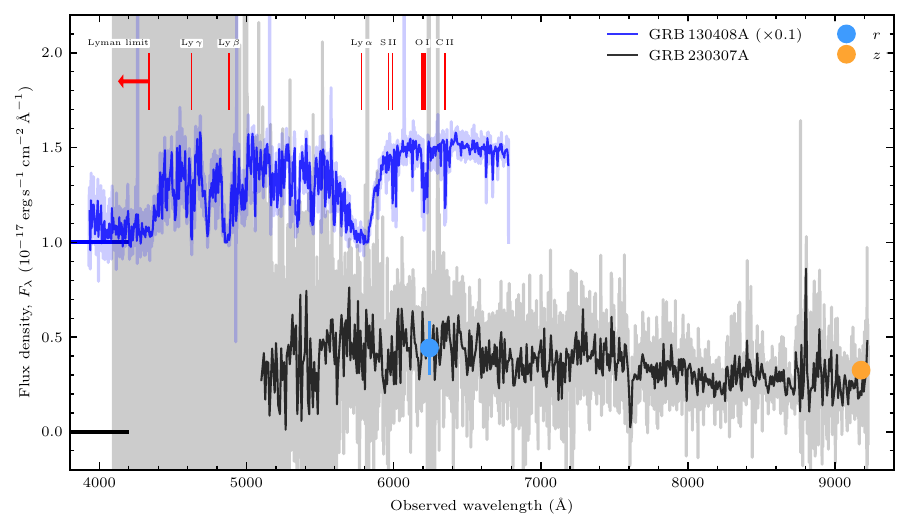}
    \caption{
        \textbf{Comparison between the Gemini spectra of GRBs~130408A and 230307A.}
        The spectrum of \GRBxx{130408A} has been scaled to better match the absolute flux level of \thisGRB. Both spectra have their original reductions plotted (faint colours), while their $3 \sigma$-clipped and $5 \times$ re-binned versions are over-plotted (solid colours). The observed photometric $rz$-band photometry obtained by Gemini and SOAR at the same epoch as the spectrum of \thisGRB\ are overlaid (from \yuhancite). A vertical offset has been applied to the \GRBxx{130408A} spectrum; the $F_\lambda = 0$ levels of both spectra are marked by horizontal lines. Many of the strong absorption features present in the \GRBxx{130408A} spectrum have been marked (vertical red bars). None of these features are present in our spectrum for \thisGRB, indicating a lower redshift.
    }
    \label{fig: GRBs 130408A + 230307A spectra}
\end{figure}

\begin{figure}
    \centering
    \includegraphics[width=\linewidth]{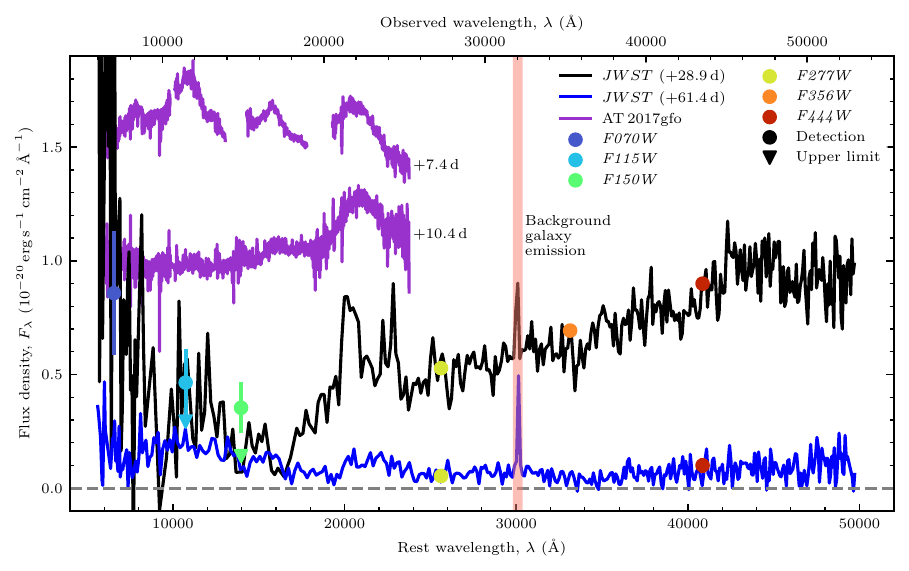}
    \caption{
        \textbf{\jwst\ spectra obtained for \thisGRB}.
        The contemporaneous photometric observations from NIRCam are also plotted, with their associated errors (note that in some cases the errors are smaller than the symbol). The location of the prominent background galaxy emission is highlighted (vertical pink band), and we mark the $F_\lambda = 0$ position with a dashed grey line. We also plot two of the late phase \xsh\ spectra (7.4 and 10.4~days post-GRB) of \gfo\ (arbitrarily scaled and offset), to illustrate the presence of a prominent spectral feature coincident with \thisGRB\ (at $\sim 2.1$\,\micron).
    }
    \label{fig: JWST spectra - Epochs 1+2}
\end{figure}

\begin{figure}
    \centering
    \begin{overpic}[width=0.65\linewidth]{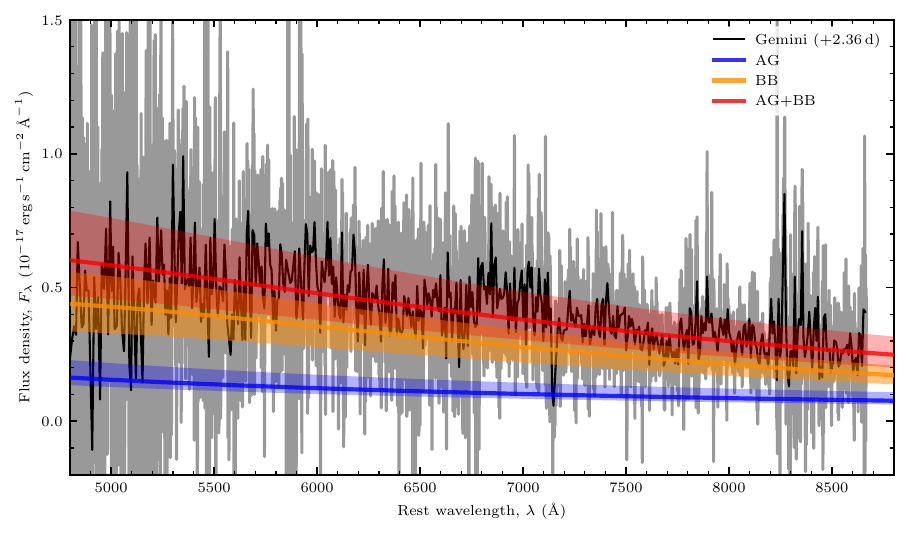}
        \put (94,49) {\footnotesize \textbf{A}}
    \end{overpic}
    \begin{overpic}[width=0.65\linewidth]{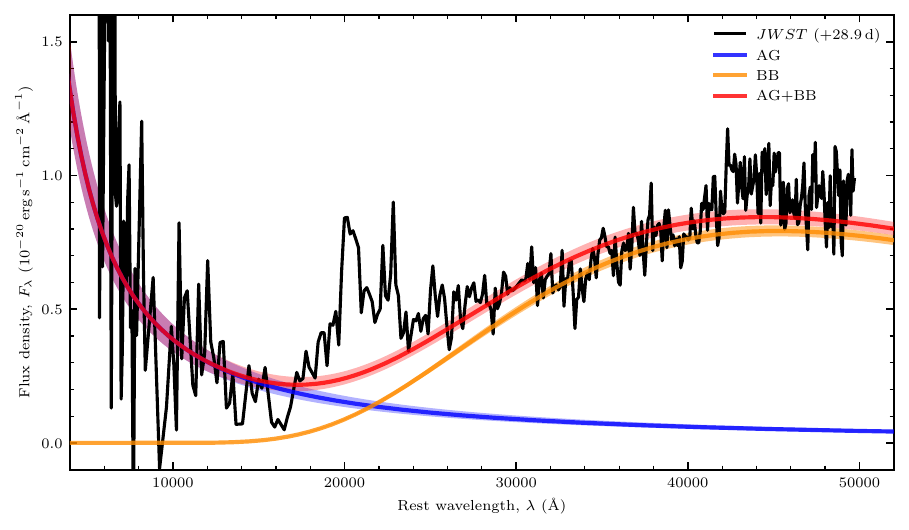}
        \put (94,49) {\footnotesize \textbf{B}}
    \end{overpic}
    \begin{overpic}[width=0.65\linewidth]{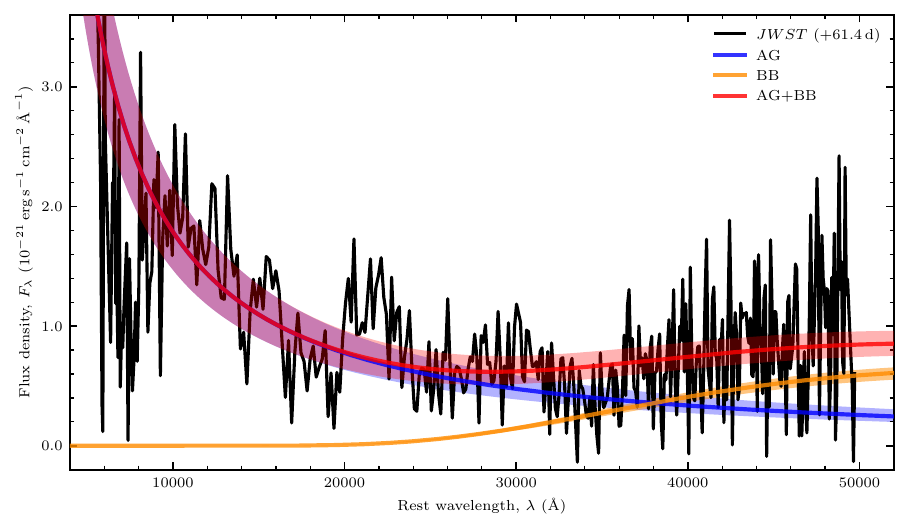}
        \put (94,49) {\footnotesize \textbf{C}}
    \end{overpic}
    \caption{
        \textbf{Comparison between our best-fit afterglow and blackbody models and the observed spectra.}
        The observed spectra have been corrected for extinction along the line of sight \cite{Schlafly2011}, and shifted to the rest frame. The best-fit afterglow (AG), blackbody (BB) and combined (AG+BB) models are plotted (blue, orange and red lines, respectively). The shaded bands correspond to $1 \sigma$ uncertainty regions.
        \textbf{Panel A:} Comparison to the $T_0 + 2.4$ day Gemini spectrum. This observed spectrum has its original reduction plotted (grey), with the $3 \sigma$-clipped and $5 \times$ re-binned version over-plotted (black).
        \textbf{Panel B:} Comparison to the $T_0 + 29$~day \jwst\ spectrum. The prominent galaxy emission line has been masked for clarity.
        \textbf{Panel C:} Comparison to the $T_0 + 61$~day \jwst\ spectrum. The prominent galaxy emission line has been masked for clarity.  
        }
    \label{fig: Spectra versus AG+BB fits}
\end{figure}

\begin{figure}
    \centering
    \includegraphics[width=\linewidth]{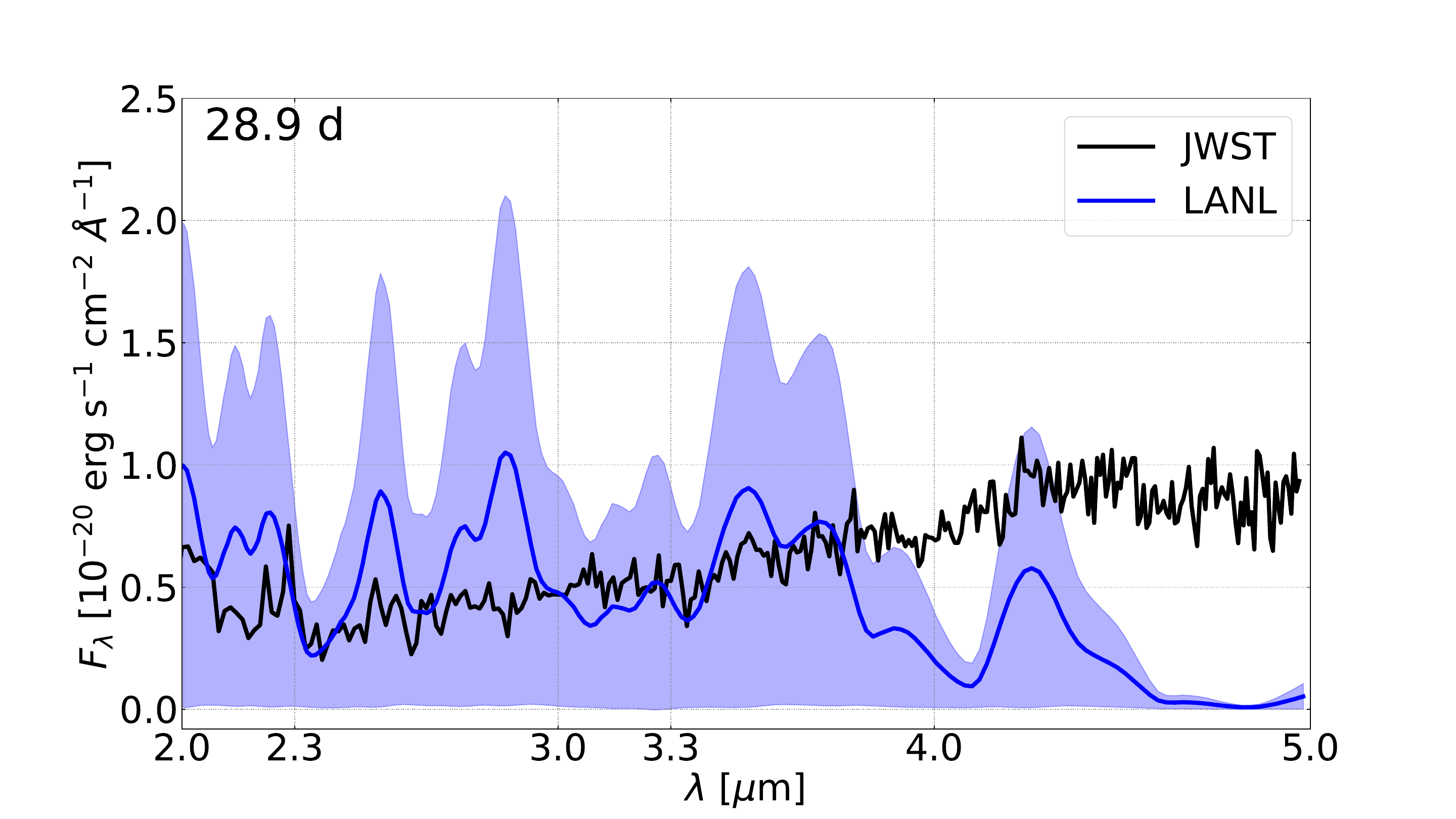}
    \caption{
        \textbf{Model spectra generated using the LANL simulation emulator (blue) compared with the afterglow-subtracted $T_0 + 29$~day \jwst\ spectrum (black).}
        The LANL model spectrum is generated using the best-fit parameters recovered from our \supernu\ analysis. The shaded region corresponds to our model $1 \sigma$ errors.
    }
    \label{fig: fit_lanl_m19_sigmaobs}
\end{figure}

\begin{figure}
    \centering
    \includegraphics[width=\linewidth]{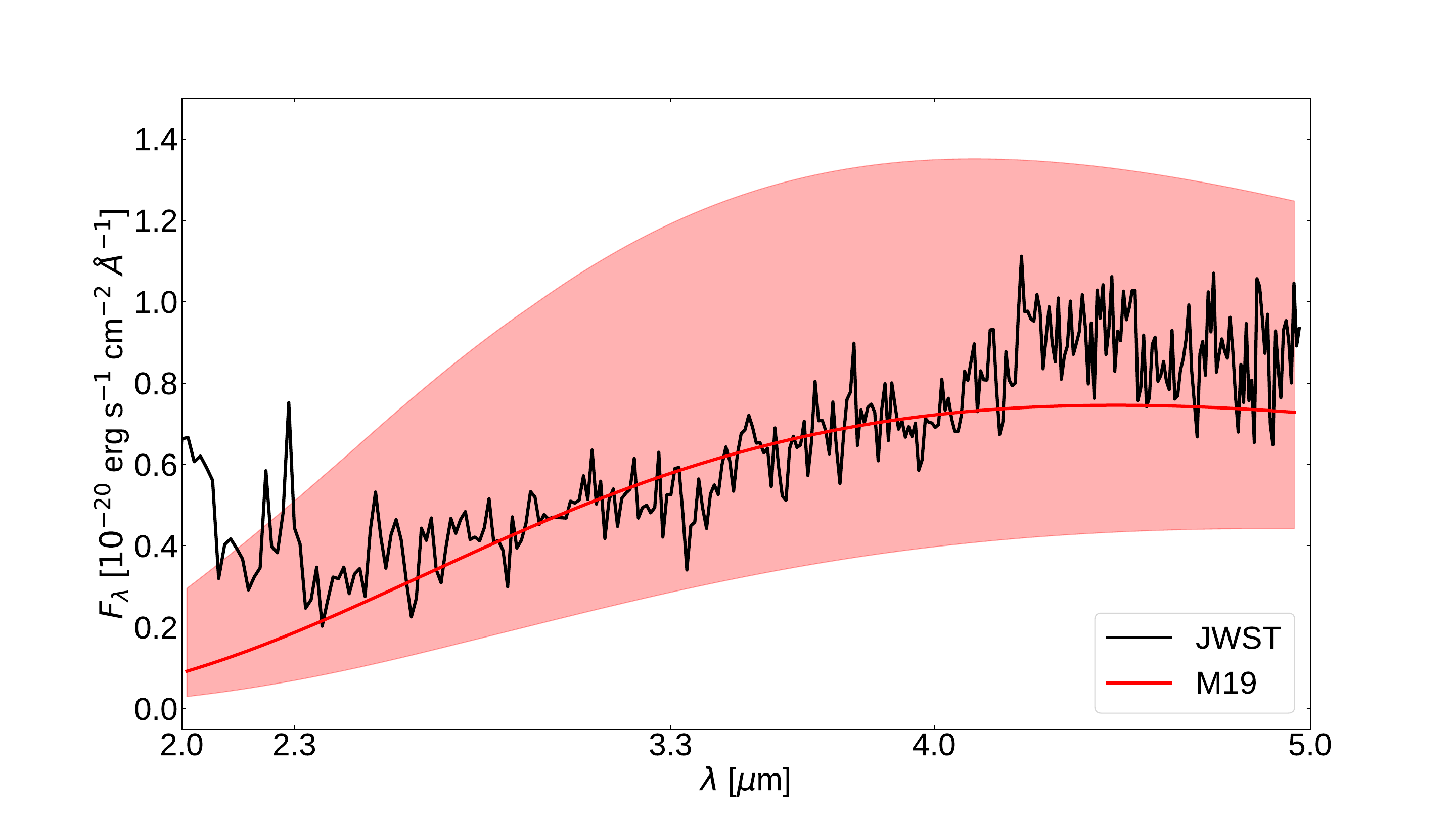}
    \caption{
        \textbf{Comparison between the $T_0 + 29$~day \jwst\ spectrum and our best-fit \metzger\ model.}
        The observed spectrum has been afterglow-subtracted (black line), and is compared to the \metzger\ best-fit model (red), generated using model parameters obtained from the posteriors in Extended Data Figure~\ref{fig: metzger_corner}. The shaded red region corresponds to the range of spectra that are generated by the $1 \sigma$ limits on the input parameters.
    }
    \label{fig: metzger_fits}
    \end{figure}

\begin{figure}
    \centering
    \includegraphics[width=\linewidth]{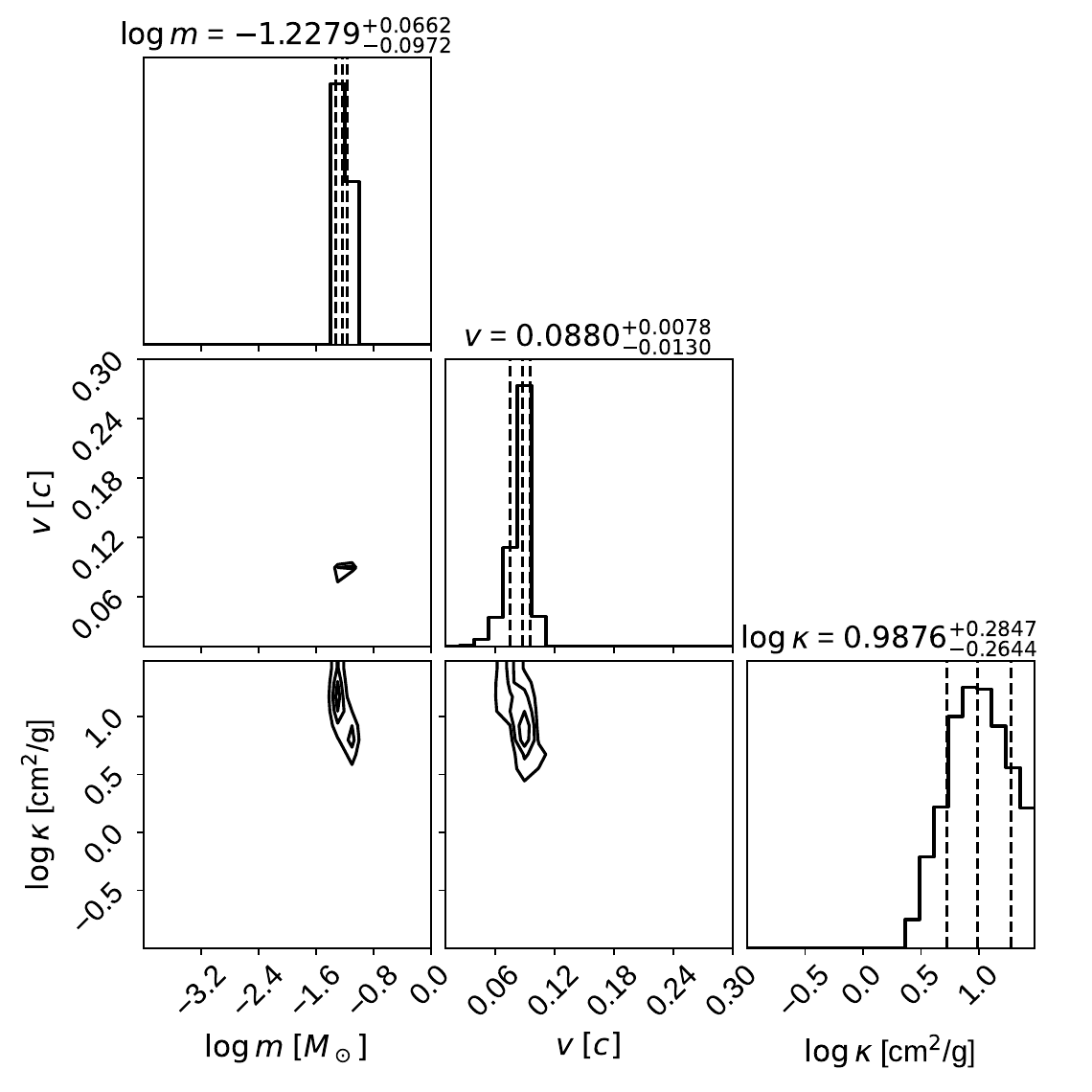}
    \caption{
        \textbf{Posterior distributions for the best-fit \metzger\ model parameters.}
        These inputs produce the best-fit model for the observed \jwst\ spectrum, presented in Extended Data Figure~\ref{fig: metzger_fits}. The \metzger\ model parameters are ejecta mass $m$, velocity $v$, and grey opacity $\kappa$. Parameter uncertainty limits reflect $1 \sigma$ confidence intervals. The velocity distribution parameter $\beta$ was treated as a free parameter and found to be unconstrained within the region $1 \leq \beta \leq 8$.
    }
    \label{fig: metzger_corner}
\end{figure}

\begin{figure}
    \centering
    \includegraphics[width=\linewidth]{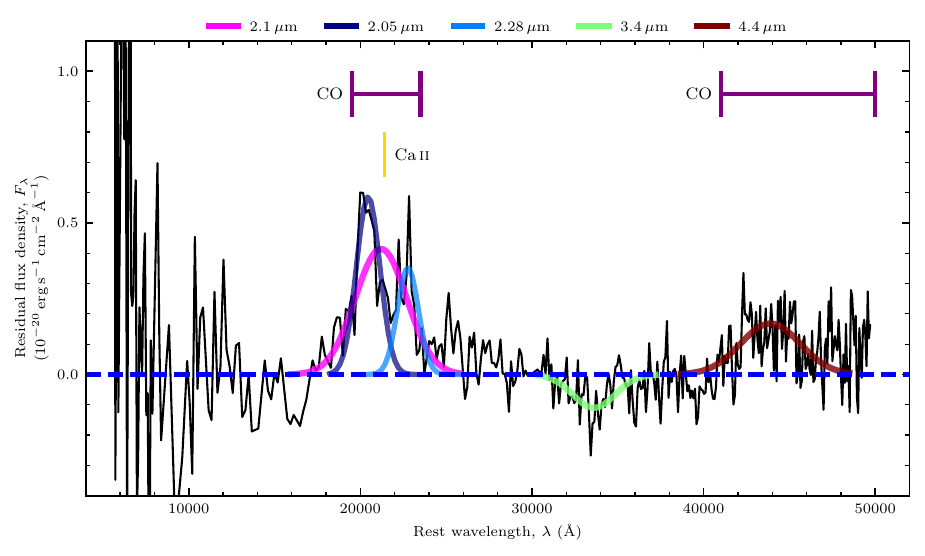}
    \caption{
        \textbf{Line identifications for the $T_0 + 29$~day \jwst\ spectral features, assuming the WD--NS/BH merger scenario.}
        Same as Figure~\ref{fig: JWST +29d spec, BB fits, continuum subtractions + Gaussian features}\blue{B}, but with the proposed sources of the emission features in the case of a WD--NS/BH merger presented. The CO first and second overtones have been blueshifted by $0.1 \, c$.
    }
    \label{fig: JWST +29d continuum subtracted spec with WD--NS/BH line IDs}
\end{figure}

\begin{figure}
    \centering
    \includegraphics[width=\linewidth]{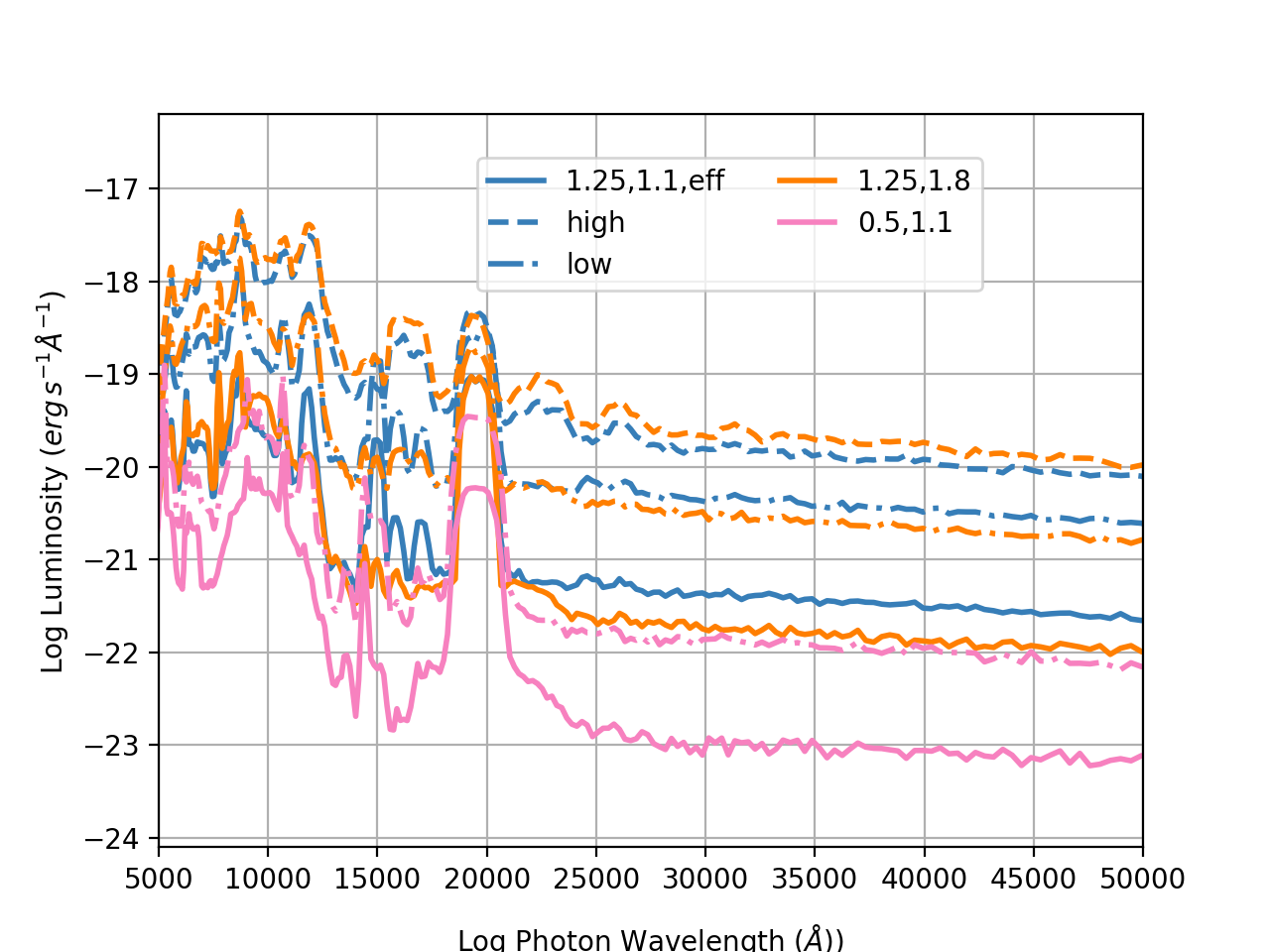}
    \caption{
        \textbf{Spectra from a series of \supernu\ models of WD--NS mergers.}
        The models include a range of white dwarf and neutron star mass pairs: $(M_{\textsc{wd}}, \, M_{{\textsc{ns}}})$\,\msun\ = (1.25, 1.1), (1.25, 1.8), and (0.5, 1.1). These masses compare the broad range of possible solutions. We also invoke a variety of accretion disk entropies based on different models of energy deposition: efficient, high and low (for more details, see \cite{Kaltenborn2022arXiv}). These models do not include non-thermal effects that will alter ionisation and excitation states (this effect may reduce the strong Ca\,\I\ feature, producing stronger Ca\,\II\ features instead, at $\sim 2.1$\,\micron). These models do not include molecular or dust opacities. 
    }
    \label{fig: WDNSmer}
\end{figure}

\clearpage
\newpage

\input{Tables/BB_fit_params}

\input{Tables/Feature_properties}

\input{Tables/M19_fit_params}

\input{Tables/M1+E2_candidate_transitions}

% Start of the Supplementary materials section
\clearpage
\newpage

\section*{Supplementary material}
% Resetting the section numbers
\setcounter{section}{0}

\section{Spectral observations} \label{sec: Methods - Spectral observations}

\thisGRB\ was observed with both Gemini and \jwst. Here we summarise the main details of each observation, and their reduction.

\subsection{Gemini}

We carried out spectroscopic observations of \thisGRB\ with the 8.1\,m Gemini South telescope through Director's Discretionary Time (DDT) program \mbox{GS-2023A-DD-104} (PI: B.~O'Connor). Observations with GMOS-S commenced on 10~March~2023 at 00:23:20 UTC ($\sim 2.36$ days post-GRB), and consisted of $4 \times 1000$\,s exposures with the R400 grating, which sampled the \mbox{$\approx 4000 - 9000$\,\AA} wavelength range. These observations were reduced using standard recipes in \pypeit\ \cite{PypeIt_paper, PypeIt_v1.12.2_zenodo}, and the spectrum was calibrated against a standard star. The reduced spectrum agrees well with the optical photometry obtained by Gemini and SOAR at the same phase \yuhancite\ (see Extended Data Figure~\ref{fig: GRBs 130408A + 230307A spectra}).

\subsection{\textit{James Webb Space Telescope}}

The \jwst\ observations of \thisGRB\ were carried out as part of DDT programs 4434 and 4445 (PI: A.~Levan) using the Near-Infrared Spectrograph \mbox{(NIRSpec; \cite{Jakobsen2022})} and the Near-Infrared Camera (NIRCam; \cite{Rieke2023}). The NIRSpec spectra were obtained with the fixed slit using the clear prism with a $\approx 1$~hour exposure. Observations for the first spectrum commenced on 5 April 2023 at 13:58:28~UTC, which corresponds to a phase of $\sim 28.9$~days post-GRB. The second spectrum was obtained almost one month later, on 8 May 2023 at 01:57:38 UTC, corresponding to $\sim 61.4$~days post-GRB.

We downloaded all calibrated and reduced Level3 data from the Mikulski Archive for Space Telescopes (MAST)\footnote{\url{https://mast.stsci.edu/portal/Mashup/Clients/Mast/Portal.html}} and performed our own detailed, customised extraction using the \jwst\ pipeline (version 1.10.2; \cite{JWST_pipeline_v1.10.2_zenodo}). We compared the results to the default extractions in MAST, and manual extractions from the \jwst\ Data Visualization Tool (Jdaviz; \cite{jdaviz_2023zndo}). We found that, in both instances, our custom reductions resulted in comparable (but slightly different) extracted spectra. The spectra obtained from the customised extractions were then flux-calibrated to the contemporaneous NIRCam photometry, to account for the lack of aperture correction and slit losses (see Extended Data Figure~\ref{fig: JWST spectra - Epochs 1+2}).

\section{Data pre-processing}

We down-sample the observational data $F_{\lambda, \rm obs}$ such that each new observational wavelength bin $\lambda_k$ contains a new observational flux value $\hat{F}_{\lambda, \text{obs}, k}$ defined as:
\begin{equation}
    \hat{F}_{\lambda, \text{obs}, k} = \frac{1}{N_k}\sum_i F_{\lambda, \text{obs}, i} \text{ for } \lambda_k \leq \lambda_i < \lambda_{k+1},
\end{equation}
where $N_k$ is the number of original observational wavelength data points $\lambda_i$ that are down-sampled into the relevant wavelength bin $\lambda_k$. We rebin the observational data to match the 1024 equally log-spaced wavelength bins spanning \mbox{$0.1 \leq \lambda \, (\mu {\rm m}) \leq 12.8$} used by \supernu, described in the following section. Throughout this work, all simulated spectra are compared to the rebinned, down-sampled observational data $\hat{F}_{\lambda, \rm obs}$. Due to the difference in wavelength ranges between our observed and simulated data sets, we are only able to compare the observed data to \emph{at most} 193 wavelength bins between $2 - 5$\,\micron. Unless otherwise noted, we down-sample our data to only include wavelengths $> 2$\,\micron\ (approximately where the thermal continuum dominates) and below 5\,\micron\ (the upper limit of the \jwst\ observation).

\section{Parameter inference}

We identify the best-fit ejecta parameters \mbox{$\vec{x}_{\texttt{M19}} = (m, v, \kappa)$} and \mbox{$\vec{x}_{\textsc{lanl}} = (m_d, v_d, m_w, v_w)$} using a simple mean-squared-error goodness-of-fit statistic, defined as:
\begin{equation}
    \chi^2 = \sum_{k = 0}^{1023} \left(\frac{F_{\lambda, \text{intp}, k} - \hat{F}_{\lambda, \text{obs}, k}}{\sigma_{\hat{F}_{\lambda, \text{obs}, k}}}\right)^2,
\end{equation}
where $k$ represents the index for wavelength bins $\lambda_k$, $F_{\lambda, \text{intp}, k}$ is the interpolated spectral energy density scaled to a distance of 291\,Mpc, $\hat{F}_{\lambda, \text{obs}, k}$ is the rebinned observed spectral energy density, and $\sigma_{\hat{F}_{\lambda, \text{obs}, k}}$ is the uncertainty on the rebinned observed spectral energy density. To assess the relative distribution of different model parameters $\vec{x}$, we use a likelihood, $\exp(\frac{-\chi^2}{2})$ and an initial uniform prior over ejecta parameters $\vec{x}$ given the limits described in the LANL and \metzger\ modelling sections. From our posterior-weighted Monte Carlo samples, we report the median posterior value as the best-fit sample, with statistical errors on each component derived from the posterior distribution.

\section{Line identification}

As highlighted above, the $T_0 + 29$~day \jwst\ spectrum contains evidence of spectral features in the ejecta at these late times. There exist two interpretations of the continuum location (\chisqall\ and \chisqred), which result in the inference of either one absorption ($\sim 3.4$\,\micron) and two emission ($\sim 2.1$ and 4.4\,\micron) features in the \chisqall\ case, or one emission ($\sim 2.1$\,\micron) and two absorption ($\sim 3.4$ and 4.0\,\micron) features in the \chisqred\ case. We investigate the cause of each of these features below. 

\subsection{Emission features}

\subsubsection{Methodology}

For emission features at these late phases, it is reasonable to expect their source to be from intrinsically weak (\ie, forbidden) lines, emitted from some optically thin region of the ejecta material \cite{Hotokezaka2021, Pognan2022a, Pognan2022b}. We therefore search for forbidden transitions (either magnetic dipole or electric quadrupole, traditionally labelled M1 and E2, respectively) that match either of the emission features inferred for \thisGRB. We utilise the same methodology outlined by \cite{Gillanders2023arXiv} to generate our synthetic line lists for all heavy elements of interest (neutral to doubly-ionised, for $Z = 38 - 92$), taking level information from the National Institute of Standards and Technology Atomic Spectra Data base (NIST ASD; \cite{NIST2022}).

We follow the same method as outlined by \cite{Gillanders2023arXiv}, and we summarise the main steps below (but for full details, see \cite{Gillanders2023arXiv}). Basically, we generate synthetic line lists from the level information presented by the NIST ASD \cite{NIST2022}, for all neutral, singly- and doubly-ionised species, with $Z = 38 - 92$. Although the radiation temperature inferred for the KN ejecta at $T_0 + 29$~days is $< 1000$\,K, fast electrons produced by radioactive decay may be capable of dominating ionisation above the photosphere \cite{Hotokezaka2021}. This effect can increase the dominant ionisation stage from $\sim$~neutral, up to $\sim$~doubly-ionised. We therefore include data for all three of these ion stages (\I--\III), although a quantitative investigation -- including both non-thermal and non-local thermal \linebreak equilibrium effects -- is necessary to determine the exact ionisation balance for the ejecta at this phase. 

With this level information, we perform a rudimentary set of calculations, whereby we compute the energy differences between all levels. This returns a list of all transitions that are theoretically possible, between all known levels in the NIST ASD. Since we are interested in forbidden M1 and E2 transitions, we apply their transition rules to subsequently filter our line lists. We apply the LS-coupling rules (where applicable), and discard any transitions that have an upper level with an energy greater than the ionisation threshold of the species under investigation. We perform a crude estimate for the relative strengths of the transitions of each ion (as in \cite{Gillanders2023arXiv}), and discard any negligibly weak transitions from our comparisons to the data (\ie, those with strengths $< 10^{-3}$ that of the strongest line within $\lambda = 1.5 - 5$\,\micron). We apply an upper level energy cut at $E_{\textsc{u}} = 3$\,eV, and discard any lines that originate from upper levels that exceed this (since we do not expect these levels to be significantly populated in a $\sim 1000$\,K plasma). Finally, we compare the line locations and strengths with the observed spectrum, and determine if there is agreement between any of the strong transitions of each ion and the observed emission feature locations, adopting a tolerance corresponding to a $0.05 \, c$ Doppler shift.

We note that the level information contained within the NIST ASD has been critically evaluated, and so our compiled line list contains transition wavelengths that are accurate enough for line identification studies. 

Currently available atomic line lists either: (i)~contain complete line lists with estimates for intrinsic line strength, but have not been calibrated, and so the wavelengths are not accurate enough for line identification studies, or (ii)~contain calibrated (and therefore accurate) transitions, but are significantly incomplete, particularly for heavy elements and for long wavelengths (which we are most interested in here). Both of these factors impact any detailed line identification study. Therefore, despite the simplicity of our generated line lists, they represent the current best method to capture transition information for many of the heavy \rpro\ elements (with some exceptions, see \eg, the Ce\,\III\ and Th\,\III\ analysis below). As shown by \cite{Gillanders2023arXiv}, this framework is able to recover the results of other line identification studies for \gfo, motivating the validity of this approach.

We use this line list to search for ions that are expected to produce prominent emission only in the regions of the observed emission features ($\sim 2.1$ and 4.4\,\micron). We present our shortlist of possible transitions in Extended Data Table~\ref{tab: Forbidden candidate transitions}, and the locations of the most promising of these are marked in Figure~\ref{fig: JWST +29d spec, BB fits, continuum subtractions + Gaussian features}. We rule out ions if there is no good agreement with the observed emission features. We also rule out any ions that match an emission feature, but possess other prominent transitions that are discrepant with the observed spectrum. Our analysis favours ions that have only a small number of strong transitions within our observable wavelength range, most (or all) of which match the feature locations of the observed data. Also, due to the low temperature invoked ($T = 1000$\,K), our analysis heavily favours transitions between the lowest-lying energy levels.

\subsubsection{Results and discussion}

We find agreement with the observed data for a number of ions, including [Ag\,\III], [Sn\,\II], [Te\,\I], [Te\,\III], [Ba\,\I], [Ba\,\II], [Er\,\I], [Er\,\II], [Er\,\III], [Hf\,\I], [W\,\III], [Ir\,\II] and [Pt\,\II] for the $\sim 2.1$\,\micron\ feature (or the $\sim$ 2.05, 2.28\,\micron\ features). For the $\sim 4.4$\,\micron\ emission feature, we shortlist [In\,\I], [Te\,\II], [Ba\,\I], [Nd\,\I], [Nd\,\III], [Hf\,\I], [W\,\III] and [Ac\,\I]. These shortlists contain ions from all three \rpro\ peaks, as well as some lanthanides, and an actinide. Further study of these species, with the aim of generating intrinsic line strengths for their transitions, are necessary to confirm the contribution of any of these transitions to the data. However, we can still reflect upon the validity of these shortlisted transitions based on other works in the literature.

Ref.~\cite{Hotokezaka2023arXiv} present an analysis of doubly-ionised tellurium forbidden emission. They demonstrate that the transition between two terms in the ground configuration of Te\,\III\ gives rise to a transition at 2.10\,\micron\ that may be responsible for the emission feature at $\sim 2.07$\,\micron\ in \gfo\ (see Extended Data Figure~\ref{fig: JWST spectra - Epochs 1+2}; \cite{Gillanders2023arXiv, Hotokezaka2023arXiv}). This [Te\,\III] transition is the same line we shortlist here (see Figure~\ref{fig: JWST +29d spec, BB fits, continuum subtractions + Gaussian features}, Extended Data Table~\ref{tab: Forbidden candidate transitions}), and is also the line that \cite{Levan2023arXiv} favour as the source of the emission feature at $\sim 2.1$\,\micron\ in their analysis of \thisGRB. The presence of this line is supported not only by its intrinsic strength, but also by nucleosynthetic arguments. As discussed in \cite{Gillanders2023arXiv}, Te lies at the top of the second \rpro\ abundance peak, and so it is expected to be produced in a wide array of \rpro\ scenarios.

In addition to the coincidence with this [Te\,\III] transition and the $\sim 2.1$\,\micron\ emission feature, we note the agreement with the strongest lines of the other ions of Te. Neutral Te possesses two transitions that are also coincident with this $\sim 2.1$\,\micron\ emission feature, while singly-ionised Te possesses a strong transition in agreement with the location of the other inferred emission peak, at $\sim 4.4$\,\micron\ (see Extended Data Table~\ref{tab: Forbidden candidate transitions}). The agreement between the strongest forbidden transitions of the lowest three ion stages of Te and the two observed emission peaks in the observed data warrants further study. A quantitative investigation of the ionisation balance of the ejecta at this phase is needed to determine the dominant ion (or ions). Some scenario where singly- and doubly-ionised Te are both present would (at least based on our rudimentary analysis here) be capable of explaining both prominent emission features in the \jwst\ observations at $T_0 + 29$~days.

Ref.~\cite{Bondarev2023} have previously investigated singly-ionised tin (Sn\,\II), and concluded that the 2.3521\,\micron\ M1 line (see Extended Data Table~\ref{tab: Forbidden candidate transitions}) is expected to be the most prominent of all [Sn\,\II] transitions across a range of plausible KN ejecta conditions. Although their calculations focussed on hotter temperatures \mbox{($T = 2500 - 4500$\,K)} than exhibited by \thisGRB, this M1 transition remains prominent across all temperatures, since it arises from a transition between two levels in the ground term. This implies that if singly-ionised tin is present in the ejecta of \thisGRB, its most prominent observational signature would be a late-time emission feature centred at 2.3521\,\micron. Tin is another second \rpro\ peak element that is expected to be abundant across a range of \rpro\ scenarios \cite{Gillanders2022}. Therefore, it seems reasonable to expect to see evidence of it in the observations, especially if we can constrain the presence of other second \rpro\ peak elements (\eg, Te).

Ref.~\cite{Gillanders2021} present detailed analysis of the presence of platinum and gold, from neutral to doubly-ionised (Pt\,\I--\III, Au\,\I--\III), in the context of kilonova ejecta. They utilise a complete atomic data set that has been calibrated to experiment, meaning they have access to a complete line list, which possesses accurate line wavelengths, in addition to estimates for the intrinsic line strengths. Crucially, this dataset possesses information for both strong permitted transitions (\ie, electric dipole, or E1, transitions) and intrinsically weak forbidden transitions (M1 and E2). In their analysis, they investigate a range of plausible late-time kilonova ejecta conditions, and in addition to comparing to the spectra of \gfo, they make a number of predictions for the presence of prominent lines that may appear in the spectra of future observations of kilonovae. Specifically, they propose that the [Pt\,\II] 2.1883\,\micron\ M1 line we shortlist here (see Extended Data Table~\ref{tab: Forbidden candidate transitions}) is the second most prominent forbidden transition expected to be produced by [Pt\,\II] across all wavelengths (the 1.1877\,\micron\ line is expected to be stronger, but it lies in the wavelength range dominated by afterglow emission, and so might not be visible in our \jwst\ spectrum).

Ref.~\cite{Gillanders2021} also predict the presence of a strong [Au\,\II] M1 transition, with \mbox{$\lambda = 3.8446$\,\micron}. This is expected to be the strongest [Au\,\II] transition above 1\,\micron, yet it is not detected in the \jwst\ spectra. While Pt and Au are expected to be co-produced (since they lie at the top of the third \rpro\ abundance peak), Au is expected to be less abundant than Pt (by a factor of $\sim$~a few; \cite{Gillanders2021}), due to Au being an odd-$Z$ element. Therefore, it seems plausible that Pt may have an observable impact on the observed spectrum at 2.1883\,\micron, without Au also having an observable effect (at 3.8446\,\micron).

Alternatively, the feature at $\sim 2.1$\,\micron\ may not possess any contribution from Pt emission, and may simply represent coincident agreement. With only a single transition matched to the data, it is impossible to definitively link the emission to [Pt\,\II]. This point applies not only to [Pt\,\II], but to all shortlisted transitions we present in Extended Data Table~\ref{tab: Forbidden candidate transitions}. Of our shortlisted species, the ones that are most convincing (and are of most interest) are those that match multiple features, or possess matches to the data across multiple ions.

[Ba\,\I], [Hf\,\I] and [W\,\III] possess multiple strong lines in agreement with the data. Similar to the Te case, elements that can reproduce the data with multiple ions are of interest, since positive identification of these would enable some inferences about the ionisation of the ejecta material to be made. Nd\,\I~\&~\III\ both possess transitions capable of producing the $\sim 4.4$\,\micron\ emission feature, while Er\,\I,~\II~\&~\III\ all possess strong transitions that match the data at $\sim 2.1$\,\micron.

Complete and accurate atomic data is sorely lacking for heavy ions, and for NIR--MIR wavelengths. Further work to improve the available atomic data is of utmost importance to interpret future \jwst\ observations of kilonovae. We propose the species presented in Extended Data Table~\ref{tab: Forbidden candidate transitions} are those capable of producing the observed emission features of \thisGRB. We conclude that these species (with particular emphasis on the ones discussed above) are the ones that most pertinently need improved atomic data, to further investigate their contribution to the observed emission of \thisGRB.

\subsection{Absorption features}

We explore elemental species that can reproduce the absorption features at \mbox{$\sim 3.4$} and 4.0\,\micron. Candidate species for strong absorption features in the spectra of the kilonova \gfo\ include He, Sr, La and Ce \cite{Watson2019, Domoto2021, Domoto2022, Gillanders2022, Perego2022, Tarumi2023arXiv}. These elements all have small numbers of valence electrons, and thus, each transition tends to be strong \cite{Domoto2022}. Also, lanthanides and actinides tend to have transitions at infrared wavelengths, due to small energy level spacings \cite{Kasen2013, Fontes2020, Tanaka2020, Fontes2023}. Because of these atomic properties, we focus on lanthanides and actinides with a small number of valence electrons as candidate elements for the absorption features in the $T_0 + 29$~day \jwst\ spectrum.

A large part of kilonova ejecta is optically thin at this epoch, and fast electrons produced by radioactive decay may dominate ionisation above the photosphere. In such conditions, typical ions of lanthanides and actinides are expected to be singly- or doubly-ionised \cite{Hotokezaka2021}. Therefore, singly- or doubly-ionised lanthanides or actinides with a small number of valence electrons can be good candidates for strong absorption features. Although atomic data of such heavy elements are not complete, in particular at infrared wavelengths, accurate atomic data has been constructed for Ce\,\III\ and Th\,\III\ \cite{Domoto2022}. Thus, here we explore the wavelengths and strengths of Ce\,\III\ and Th\,\III\ lines by extending the line list of \cite{Domoto2022}, up to 5\,\micron.

Figure~\ref{fig: JWST +29d spec, BB fits, continuum subtractions + Gaussian features} shows the positions of the strongest Ce\,\III\ and Th\,\III\ lines, blueshifted by $v = 0.1 \, c$, and compared with the continuum-subtracted $T_0 + 29$~day \jwst\ spectrum. We only show the lines that we expect to be strong; \ie, those with $\log (gf) \geq -2$ and originating from low-lying levels ($E < 10^4$\,cm$^{-1}$), that produce a large Sobolev optical depth. For the evaluation of the Sobolev optical depth, we adopt the abundance patterns to match with the \rpro\ residuals of the solar abundances at $A > 88$ (`Solar' model in \cite{Domoto2021}). For the ionisation, we do not take non-thermal ionisation into account, but instead solve the Saha equation for a density, $\rho = 10^{-16}$\,\gcm, as used for the opacity study, and a temperature of $T = 5000$\,K, which is typical for kilonova nebular ejecta \cite{Hotokezaka2021}.

We find that the absorption feature at $\sim 3.4$\,\micron\ can be matched with a strong Ce\,\III\ line, but this requires a velocity, $v \approx 0.02 \, c$, lower than the inferred photospheric velocity. Interestingly, we also expect strong absorption features around $\sim 1.5$\,\micron, as proposed for the NIR feature of \gfo\ \cite{Domoto2022, Tanaka2023arXiv, Gillanders2023arXiv}. However, the signal-to-noise ratio around this wavelength range is not as good as at the longer wavelengths, and so no definitive statement as to the presence (or absence) of this feature can be made.

The absorption feature around $\sim 4.0$\,\micron\ may be matched with a Th\,\III\ line (see Figure~\ref{fig: JWST +29d spec, BB fits, continuum subtractions + Gaussian features}). If this feature is attributed to Th\,\III, we also expect to see several other absorption features at shorter wavelengths, which are not clearly visible in the \jwst\ spectrum. Therefore, the positive identification of Th\,\III\ is not very robust.

Although neither Ce\,\III\ or Th\,\III\ are capable of convincingly reproducing the data, the general trend of many strong transitions spanning the NIR and MIR are typical of the lanthanide and actinide elements. With more complete line information, future studies of this (and other) kilonova events will be able to better interpret the transformative observations of the NIR and MIR, only possible through the use of \jwst.

% Final journal-mandated bits
\clearpage
\newpage

\backmatter

\bmhead{Author contribution statements}
~ \\
\noindent
JHG initiated and led the project, reduced all spectral observations, identified the presence of the spectral features in the $T_0 + 29$ and $T_0 + 61$~day \jwst\ spectra, led the spectral analysis of the distance constraint, performed the combined afterglow and blackbody fitting, led the investigation of the emission features in the \jwst\ data, led the writing and editing of the manuscript, and produced Figures~\ref{fig: Gemini vs KNe and GRB-SNe}~and~\ref{fig: JWST +29d spec, BB fits, continuum subtractions + Gaussian features}, and Extended Data Figures~\ref{fig: GRBs 130408A + 230307A spectra},~\ref{fig: JWST spectra - Epochs 1+2},~\ref{fig: Spectra versus AG+BB fits}~and~\ref{fig: JWST +29d continuum subtracted spec with WD--NS/BH line IDs}.
ET helped lead the project, assisted with the interpretation of all analyses, and contributed significantly to the writing and editing of the manuscript.
CLF led the analysis and discussion of the WD--NS/BH merger scenario, led the investigation of the opacities, made Figure~\ref{fig: opacfin} and Extended Data Figure~\ref{fig: WDNSmer}, and contributed significantly to the writing of the manuscript.
MR led the \metzger\ and LANL kilonova modelling and associated parameter inference, made Extended Data Figures~\ref{fig: fit_lanl_m19_sigmaobs},~\ref{fig: metzger_fits}~and~\ref{fig: metzger_corner}, and contributed significantly to the writing and editing of the manuscript.
BO assisted with data reduction, interpretation of the evolution of \thisGRB, and contributed significantly to the editing of the manuscript.
CJF assisted with the opacity analysis and discussion.
Y-HY assisted with the empirical afterglow and blackbody fitting.
ND, SR, and MT led the investigation of the absorption features in the \jwst\ data, assisted with the interpretation of the emission features, and helped make Figure~\ref{fig: JWST +29d spec, BB fits, continuum subtractions + Gaussian features}.
ODF assisted with the detailed extraction of the \jwst\ spectra.
SD assisted with data reduction and interpretation of the evolution of \thisGRB.
All authors contributed to the writing and editing of the manuscript.

\bmhead{Acknowledgments}
~ \\
\noindent
% Gemini data processing
We thank Kathleen Labrie for useful discussions on Gemini data and \dragons, and Peter Hoeflich for his comments on the spectral features. 
% ERC
This work was supported by the European Research Council through the Consolidator grant BHianca (grant agreement ID 101002761). 
% Aspen
This work was in part carried out at the Aspen Center for Physics, which is supported by National Science Foundation grant PHY-2210452. 
% Others...

\bmhead{Data Availability}
~ \\
\noindent
All data presented in this manuscript are available upon reasonable request to the corresponding author.

\clearpage
\newpage

\bibliography{references}
% \bibliography{sn-bibliography}% common bib file

\end{document}

%% file: _commands.tex
\newcommand{\yuhancite}{\cite{Yang_partner_paper}}

\newcommand{\GRBxx}[1]{\mbox{GRB\,#1}}
\newcommand{\thisGRB}{\GRBxx{230307A}}

\newcommand{\chisqall}{\mbox{$\chi^2_{1.0+}$}}
\newcommand{\chisqred}{\mbox{$\chi^2_{4.2+}$}}

\newcommand{\jwst}{\mbox{\textit{JWST}}}
\newcommand{\xsh}{\mbox{X-shooter}}

\newcommand{\gfo}{\mbox{AT\,2017gfo}}

\newcommand{\pypeit}{\mbox{\texttt{pypeit}}}
\newcommand{\dragons}{\mbox{\texttt{DRAGONS}}}
\newcommand{\metzger}{\mbox{\texttt{M19}}}
\newcommand{\supernu}{\mbox{\texttt{SuperNu}}}

\newcommand{\eg}{\mbox{e.g.}}
\newcommand{\ie}{\mbox{i.e.}}

\newcommand{\I}{{\sc i}}
\newcommand{\II}{{\sc ii}}
\newcommand{\III}{{\sc iii}}

\newcommand{\blue}[1]{{\textcolor{blue}{#1}}}

\newcommand{\lyalpha}{\mbox{Lyman-$\alpha$}}
\newcommand{\lybeta}{\mbox{Lyman-$\beta$}}

\newcommand{\micron}{$\mu$m}
\newcommand{\rpro}{\mbox{$r$-process}}

\newcommand{\degree}{\mbox{$^\circ$}}
\newcommand{\msun}{\mbox{M$_{\odot}$}}

\newcommand{\cmg}{$\mathrm{cm}^{2}\,\mathrm{g}^{-1}$}
\newcommand{\gcm}{g\,$\mathrm{cm}^{-3}$}
\newcommand{\kmsMpc}{$\mathrm{km\,s}^{-1}\,\mathrm{Mpc}^{-1}$}
\newcommand{\ergs}{erg\,s$^{\mathrm{-1}}$}
\newcommand{\ergscmA}{erg\,s$^{-1}$\,cm$^{-2}$\,\AA$^{-1}$}

%% file: _packages.tex
\usepackage{multirow}
\usepackage{amsmath,amssymb,amsfonts}
\usepackage{xcolor}
\usepackage{manyfoot}
\usepackage{booktabs}
\usepackage{changepage}
\usepackage[percent]{overpic}
\usepackage{xfrac}
\usepackage{subcaption}

%% file: Tables/BB_fit_params.tex
\begin{table*}
    \renewcommand*{\arraystretch}{1.2}
    \centering
    \caption{
        \textbf{Ejecta parameters inferred from our Planck modelling of the observed spectra.}
        Uncertainties are quoted to $1 \sigma$.
    }
    \begin{threeparttable}
        \centering
    \begin{tabular}{lcccc}
        \toprule
        Time since      &$T_{\rm ph}$    &$R_{\rm ph}$       &$v_{\rm ph}$    &Luminosity    \\
        $T_0$ (days)    &(K)             &($10^{15}$\,cm)    &($c$)           &(\ergs)       \\
        \midrule
        2.36                &$6700 \pm 400$     &$0.45 \pm 0.05$    &$0.073 \pm 0.008$      &$3.0 \pm 0.6 \times 10^{41}$      \\
        28.9\tnote{a}       &$640 \pm 18$       &$6.79 \pm 0.16$    &$0.091 \pm 0.002$      &$5.5 \pm 0.3 \times 10^{39}$      \\
        28.9\tnote{b}       &$608 \pm 33$       &$8.29 \pm 0.78$    &$0.111 \pm 0.010$      &$6.7 \pm 1.3 \times 10^{39}$      \\
        61.4\tnote{c}       &$486 \pm 10$       &$3.83 \pm 0.01$    &$0.024 \pm 0.001$      &$5.8 \pm 0.3 \times 10^{38}$      \\
        \bottomrule    
    \end{tabular}
        \begin{tablenotes}
            \small
            \item[a] \chisqall\ best-fitting model.
            \item[b] \chisqred\ best-fitting model.
            \item[c] The validity of invoking a thermal component to model the continuum at this epoch is uncertain.
        \end{tablenotes}
    \end{threeparttable}
    \label{tab: BB temperatures and velocities}
\end{table*}

%% file: Tables/Feature_properties.tex
\begin{table}
    \renewcommand*{\arraystretch}{1.2}
    \centering
    \caption{
        \textbf{Properties inferred from modelling the \jwst\ spectral features.}
    }
    \begin{tabular}{lccc}
        \toprule
        Approx. feature            &Peak                &$v_{\textsc{fwhm}}$        &Feature flux                                 \\
        position (\micron)         &wavelength (\AA)    &($c$)                      &($10^{-18}$\,erg\,s$^{-1}$\,cm$^{-2}$)       \\
        \midrule
        \multicolumn{4}{c}{$T_0 + 29$~days (\chisqall)}    \\
        \midrule
        2.1        &21200       &0.167       &15.6       \\
        2.05       &20500       &0.079       &10.1       \\
        2.28       &22800       &0.064       &5.4        \\
        4.4        &43900       &0.093       &7.3        \\
        \midrule
        \multicolumn{4}{c}{$T_0 + 29$~days (\chisqred)}    \\
        \midrule
        3.4        &33600       &0.086       &4.2        \\
        4.0        &39700       &0.057       &4.1        \\
        \midrule
        \multicolumn{4}{c}{$T_0 + 61$~days}    \\
        \midrule
        2.1        &22000       &0.173       &3.5        \\
        \bottomrule    
    \end{tabular}
    \label{tab: Gaussian velocities}
\end{table}

%% file: Tables/M19_fit_params.tex
\begin{table}
    \normalsize
    \renewcommand*{\arraystretch}{1.2}
    \centering
    \caption{
        \textbf{Ejecta and photosphere properties inferred from our \metzger\ modelling of the spectral features in the \mbox{$T_0 + 29$~day} \jwst\ spectrum.}
        Uncertainties are quoted to $1 \sigma$.
    }
    \begin{tabular}{lcr}
        \toprule
        Parameter       &Value      &Unit      \\
        \midrule
        $m_{\rm{ej}}$       &$0.059^{+0.010}_{-0.012}$      &{[$M_\odot$]}          \\
        $v_{\rm{ej}}$       &$0.088^{+0.008}_{-0.013}$      &{[$c$]}                \\
        $\kappa$            &$10^{+9}_{-5}$                 &{[\cmg]}               \\
        $\beta$             &$4^{+3}_{-2}$                  &--                     \\
        $T_{\rm{ph}}$       &$608^{+86}_{-84}$              &{[K]}                  \\
        $R_{\rm{ph}}$       &$7.507^{+4.020}_{-1.746}$      &{[$10^{15}$\,cm]}      \\
        $L_{\rm{tot}}$      &$5.491^{+3.719}_{-1.922}$      &{[$10^{39}$\,\ergs]}   \\
        $\rho$              &$6.642^{+8.832}_{-4.576}$      &{[$10^{-17}$\,\gcm]}   \\
        
        \bottomrule    
    \end{tabular}
    \label{tab: metzger params}
\end{table}

%% file: Tables/M1+E2_candidate_transitions.tex
\begin{table*}
\begin{adjustwidth}{-1.5cm}{-2cm} % widens the permitted size of the table -- helps to fit the table on the page!
    \renewcommand*{\arraystretch}{1.2}
    \footnotesize
    \centering
    \caption{
        \textbf{Candidate forbidden (\ie, M1 and E2) transitions for the emission features in the $T_0 + 29$~day \jwst\ spectrum of \thisGRB.}
        The species, wavelength, level energies, quantum J numbers, and the relative level population of the upper level for each transition are presented. The configurations and terms are displayed (from the NIST ASD; \cite{NIST2022}), when they are easily expressible in the LS-coupling scheme.
    }
    \begin{threeparttable}
        \centering
        \begin{tabular}{lccccrlccrlcc}
        \toprule
        \multirow{2.5}{*}{Species}    &\multirow{2.5}{*}{$\lambda_{\rm vac}$ (\AA)}    &\multicolumn{2}{c}{Level energies (eV)}    &    &\multicolumn{3}{c}{Lower level}    &    &\multicolumn{3}{c}{Upper level}    &\multirow{2.5}{*}{\begin{tabular}[c]{@{}c@{}} Relative level \\ population \end{tabular}}    \\
        \cmidrule{3-4}
        \cmidrule{6-8}
        \cmidrule{10-12}
            &       &Lower      &Upper      &       &Configuration      &Term       &J      &    &Configuration      &Term      &J      &    \\
        \midrule
        \multicolumn{13}{c}{$\sim 2.1$\,\micron\ (or $\sim 2.05$ and 2.28\,\micron) feature}    \\
        \midrule \relax
        [$^{47}$Ag\,\III] 		 &21696 		 &0.000 		 &0.571 	&	 &4d$^9$ 		     &$^2$D              	     &\sfrac{5}{2} 	   &	     &4d$^9$ 		     &$^2$D 	            	 &\sfrac{3}{2} 		 &1.0 			 \\ \relax
        [$^{50}$Sn\,\II] 		 &23521 		 &0.000 		 &0.527 	&	 &5s$^2$5p 		     &$^2$P$^{\textsc{o}}$       &\sfrac{1}{2}     &	     &5s$^2$5p 		     &$^2$P$^{\textsc{o}}$		 &\sfrac{3}{2} 		 &1.0 			 \\ \relax
        [$^{52}$Te\,\I] 	     &21049 		 &0.000 		 &0.589 	&	 &5p$^4$ 		     &$^3$P                      &2 	           &	     &5p$^4$ 		     &$^3$P 	               	 &1 		         &1.0 			 \\ \relax
        [$^{52}$Te\,\I] 		 &21247 		 &0.000 		 &0.584 	&	 &5p$^4$ 		     &$^3$P            		     &2 	           &	     &5p$^4$ 		     &$^3$P              		 &0 		         &0.36 			 \\ \relax
        [$^{52}$Te\,\III] 		 &21050 		 &0.000 		 &0.589 	&	 &5s$^2$5p$^2$ 	     &$^3$P              	     &0                &	     &5s$^2$5p$^2$ 	     &$^3$P               		 &1 		         &1.0 			 \\ \relax
        [$^{56}$Ba\,\I] 		 &19944 		 &2.239 		 &2.861 	&	 &6s6p 		         &$^1$P$^{\textsc{o}}$ 	     &1                &	     &5d6p 		         &$^1$D$^{\textsc{o}}$		 &2 		         &1.0 			 \\ \relax
        [$^{56}$Ba\,\II] 		 &20518 		 &0.000 		 &0.604 	&	 &6s 		         &$^2$S              	     &\sfrac{1}{2}     &	     &5d 		         &$^2$D               		 &\sfrac{3}{2} 		 &1.0 			 \\ \relax
        [$^{68}$Er\,\I] 		 &19860 		 &0.000 		 &0.624 	&	 &4f$^{12}$6s$^2$ 	 &$^3$H                    	 &6 	           &    	 &4f$^{12}$6s$^2$ 	 &$^3$F 	           	     &4 		         &1.0 			 \\ \relax
        [$^{68}$Er\,\II] 		 &21312 		 &0.055 		 &0.636 	&	 &--                 &-- 	                     &\sfrac{11}{2}    &    	 &--                 &-- 		                 &\sfrac{9}{2} 		 &1.0 			 \\ \relax
        [$^{68}$Er\,\II] 		 &19483 		 &0.000 		 &0.636 	&	 &--                 &-- 	                     &\sfrac{13}{2}    &    	 &--                 &-- 		                 &\sfrac{9}{2} 		 &1.0 			 \\ \relax
        [$^{68}$Er\,\II] 		 &20148 		 &0.055 		 &0.670 	&	 &--                 &-- 	                     &\sfrac{11}{2}    &    	 &--                 &-- 		                 &\sfrac{7}{2} 		 &0.54 			 \\ \relax
        [$^{68}$Er\,\III] 		 &19678 		 &0.000 		 &0.630 	&	 &4f$^{12}$ 		 &$^3$H                      &6 	           &    	 &4f$^{12}$ 		 &$^3$F 	           	     &4 	         	 &1.0 			 \\ \relax
        [$^{72}$Hf\,\I] 		 &21893 		 &0.000 		 &0.566 	&	 &5d$^2$6s$^2$ 		 &$^3$F               		 &2 	           &    	 &5d$^2$6s$^2$ 		 &$^3$F 	           	     &4 	         	 &0.053 		 \\ \relax
        [$^{74}$W\,\III] 		 &22416 		 &0.000 		 &0.553 	&	 &5d$^4$ 	         &$^5$D                		 &0 	           &    	 &5d$^4$    		 &$^5$D 	           	     &2 	         	 &0.070 		 \\ \relax
        [$^{77}$Ir\,\II] 		 &20886 		 &0.000 		 &0.594 	&	 &5d$^7$($^4$F)6s    &$^5$F               		 &5 	           &    	 &5d$^7$($^4$F)6s	 &$^5$F 	           	     &4           	     &1.0 			 \\ \relax
        [$^{78}$Pt\,\II] 		 &21883 		 &0.593 		 &1.160 	&	 &5d$^8$6s 		     &$^4$F              		 &\sfrac{9}{2} 	   &    	 &5d$^8$6s    		 &$^4$F 	         	     &\sfrac{7}{2} 	     &1.0 			 \\
        \midrule
        \multicolumn{13}{c}{$\sim 4.4$\,\micron\ feature}    \\
        \midrule \relax
        [$^{49}$In\,\I] 		 &45196 		 &0.000 		 &0.274 	&	 &5s$^2$5p 		     &$^2$P$^{\textsc{o}}$ 		 &\sfrac{1}{2}  	&	     &5s$^2$5p 	      	 &$^2$P$^{\textsc{o}}$ 		 &\sfrac{3}{2}   	 &1.0 			 \\ \relax
        [$^{52}$Te\,\II] 		 &45466 		 &1.267 		 &1.540 	&	 &5s$^2$5p$^3$ 	     &$^2$D$^{\textsc{o}}$ 		 &\sfrac{3}{2}   	&	     &5s$^2$5p$^3$	     &$^2$D$^{\textsc{o}}$ 		 &\sfrac{5}{2}   	 &1.0 			 \\ \relax
        [$^{56}$Ba\,\I] 		 &44847 		 &2.635 		 &2.911 	&	 &5d$^2$ 		     &$^3$F              		 &3               	&	     &5d$^2$ 	    	 &$^3$P                		 &1             	 &0.33 			 \\ \relax
        [$^{56}$Ba\,\I] 		 &43570 		 &2.681 		 &2.966 	&	 &5d$^2$ 		     &$^3$F              		 &4               	&	     &5d$^2$ 	       	 &$^3$P                 	 &2             	 &0.30 			 \\ \relax
        [$^{56}$Ba\,\I] 		 &43956 		 &2.595 		 &2.878 	&	 &5d$^2$ 		     &$^3$F              		 &2               	&	     &5d$^2$      		 &$^3$P               		 &0             	 &0.16 			 \\ \relax
        [$^{60}$Nd\,\I] 		 &42255 		 &0.000 		 &0.293 	&	 &4f$^4$6s$^2$ 	     &$^5$I 	               	 &4                	&	     &4f$^4$6s$^2$ 		 &$^5$I                 	 &6             	 &1.0 			 \\ \relax
        [$^{60}$Nd\,\III] 		 &41884 		 &0.000 		 &0.296 	&	 &4f$^4$ 		     &$^5$I 	              	 &4                	&	     &4f$^4$ 		     &$^5$I                 	 &6             	 &1.0 			 \\ \relax
        [$^{72}$Hf\,\I] 		 &42433 		 &0.000 		 &0.292 	&	 &5d$^2$6s$^2$ 	     &$^3$F 	               	 &2                	&	     &5d$^2$6s$^2$ 		 &$^3$F                 	 &3             	 &1.0 			 \\ \relax
        [$^{72}$Hf\,\I] 		 &45229 		 &0.292 		 &0.566 	&	 &5d$^2$6s$^2$ 	     &$^3$F 	               	 &3                	&	     &5d$^2$6s$^2$ 		 &$^3$F                 	 &4             	 &0.053 		 \\ \relax
        [$^{74}$W\,\III] 		 &44322 		 &0.000 		 &0.280 	&	 &5d$^4$ 		     &$^5$D 	               	 &0                	&	     &5d$^4$ 	     	 &$^5$D                 	 &1             	 &1.0 			 \\ \relax
        [$^{74}$W\,\III] 		 &45352 		 &0.280 		 &0.553 	&	 &5d$^4$ 		     &$^5$D 	               	 &1                	&	     &5d$^4$ 	     	 &$^5$D                 	 &2             	 &0.070			 \\ \relax
        [$^{89}$Ac\,\I] 		 &44814 		 &0.000 		 &0.277 	&	 &6d7s$^2$ 		     &$^2$D 	              	 &\sfrac{3}{2}   	&	     &6d7s$^2$     		 &$^2$D                      &\sfrac{5}{2}  	 &1.0 			 \\ 
        \bottomrule
        \end{tabular}
        \begin{tablenotes}
            \footnotesize
            \item[o] Denotes an odd parity.
        \end{tablenotes}
    \end{threeparttable}
    \label{tab: Forbidden candidate transitions}
\end{adjustwidth}
\end{table*}